\documentclass[aps,prd, notitlepage, onecolumn,superscriptaddress,floatfix,letterpaper,nofootinbib, longbibliography]{revtex4-1}
\usepackage{amsmath, amsthm, amssymb,slashed}

\usepackage[usenames, dvipsnames]{color}
\usepackage[svgnames]{xcolor}
\usepackage[colorlinks,citecolor=RoyalBlue, urlcolor=RoyalBlue, linkcolor=RoyalBlue ]{hyperref} %BlueViolet  NavyBlue RoyalBlue MidnightBlue

\usepackage[normalem]{ulem}

\usepackage{booktabs}

\definecolor{mygray}{gray}{0.6}

\usepackage{upgreek}
\usepackage{bbm}

\newenvironment{myfont}[2][]{\csname#2\endcsname[#1]}{}

\usepackage{slashed}
\usepackage[makeroom]{cancel}
\usepackage[normalem]{ulem}
\usepackage{soul}
\newcommand{\stkout}[1]{\ifmmode\text{\sout{\ensuremath{#1}}}\else\sout{#1}\fi}

\usepackage{sseq}
\usepackage[all,cmtip]{xy}
\usepackage{tikz-cd}
\usepackage{tikz}
\usetikzlibrary{matrix}
\usetikzlibrary{decorations.markings}
\usetikzlibrary{tikzmark,decorations.pathreplacing,positioning}
\usepackage{amsfonts}

\newcommand{\bea}{\begin{eqnarray}}
\newcommand{\eea}{\end{eqnarray}}
\def\be{\begin{equation}}
\def\ee{\end{equation}}

\newcommand{\ii}{\hspace{1pt}\mathrm{i}\hspace{1pt}}

\definecolor{red}{rgb}{1,0,0}
\definecolor{blue}{rgb}{0,0,1}
\definecolor{dblue}{rgb}{0,0,0.4}
\definecolor{green}{rgb}{0,1,0}
\definecolor{black}{rgb}{0,0,0}
\definecolor{white}{rgb}{1,1,1}

\definecolor{brn}{rgb}{.8,.4,.0}
\definecolor{redo}{rgb}{1,.5,.0}
\definecolor{ddgrn}{rgb}{0,0.4,0}
\definecolor{dgrn}{rgb}{0,0.55,0}
\definecolor{dbl}{rgb}{0,0,0.5}

\usepackage[bbgreekl]{mathbbol}
\usepackage{amscd}

\newcommand{\Z}{\mathbb{Z}}

\newcommand{\R}{\mathbb{R}}

\newcommand{\dd}{\mathrm{d}}

\newcommand{\eq}[1]{eq.~(\ref{#1})} 
\newcommand{\eqq}[1]{(\ref{#1})}

\newcommand{\bpm}{\begin{pmatrix}}
\newcommand{\epm}{\end{pmatrix}}
\newcommand{\bmm}{\begin{matrix}}
\newcommand{\emm}{\end{matrix}}

\newcommand{\cA}{ {\cal A} }

\newcommand{\cG}{ {\cal G} }

\def\Z{{\mathbb{Z}}}

\def\R{{\mathbb{R}}}

\def \Hom{\operatorname{Hom}}

\def \H{\operatorname{H}}

\def \Z{\mathbb{Z}}

\def\Ext{\operatorname{Ext}}
\newcommand {\emptycomment}[1]{}

\def\TP{\mathrm{TP}}
\def\Sq{\mathrm{Sq}}
\def\B{\mathrm{B}}

\newcommand{\SO}{{\rm SO}}
\newcommand{\Spin}{{\rm Spin}}
\newcommand{\U}{{\rm U}}
\newcommand{\SU}{{\rm SU}}

\usepackage{centernot}
\newcommand{\nn}{{\nonumber}}
\newcommand{\Sec}[1]{Sec.~\ref{#1}} 
\newcommand{\App}[1]{App.~\ref{#1}} 

\usepackage{enumitem} %{enumerate}
\usepackage{mathtools,amssymb,varwidth}

\newcommand{\Fig}[1]{Fig.~\ref{#1}} 
\newcommand{\Table}[1]{Table \ref{#1}}

\usepackage{datetime}

\newcommand{\rF}{{\rm F}}

\newcommand{\SM}{{\rm SM}}

\usepackage{multirow}

\begin{document}

%\begin{titlepage}

\title{Topological Responses of the Standard Model Gauge Group}

\author{Zheyan Wan}
\email{wanzheyan@bimsa.cn}
\affiliation{Beijing Institute of Mathematical Sciences and Applications, Beijing 101408, China}

\author{Juven Wang}
%{\includegraphics[height=3.35ex]{1280px-Wang}${}^{\ccblue{\text{W}^+}}$}%}
\email{jw@lims.ac.uk}
%\email{jw@cmsa.fas.harvard.edu}
%\href{http://sns.ias.edu/~juven/}{http://sns.ias.edu/$\sim$juven/} 
%\homepage{http://sns.ias.edu/~juven/}
%\homepage{http://idear.info/}
\affiliation{London Institute for Mathematical Sciences, Royal Institution, W1S 4BS, UK}
\affiliation{Center of Mathematical Sciences and Applications, Harvard University, MA 02138, USA}

\author{Yi-Zhuang You}
\email{yzyou@physics.ucsd.edu}
%\thanks{Dedicated to Professor Shing-Tung Yau \includegraphics[scale=0.2]{ST-Yau.PNG} and his 76th Birthday celebration on April 4th, 2025.}
\affiliation{Department of Physics, University of California, San Diego, CA 92093, USA}

\begin{abstract} 
%\today, \currenttime
While the local Lie algebra of the Standard Model (SM) of particle physics is
known as $su(3) \times  su(2) \times u(1)$,
thus far the global Lie group 
$G_{{\rm SM}_{\rm q}}=({\SU(3) \times   \SU(2) \times \U(1)})/{\Z_{\rm q}}$ with ${\rm q}=1,2,3,6$, is indecisive theoretically and experimentally.
Built upon the previous work on the 4d anomalies and 5d cobordism invariants (namely, 5d invertible field theory [iFT] or symmetry-protected topological state [SPTs]) of the SM gauge group, we further enumerate lower-dimensional iFT / SPTs in 4d, 3d, 2d, and 1d.
While the 4d SPTs are gapped phases attachable to the SM, those integer classes of SPTs (either the torsion or the free cobordism class) are not universal and are difficult to detect. However, fractional SPTs are more universal and can be detected by a topological response similar to the Hall conductance under an appropriate symmetry twist background field.
Recently the SM's {\bf symmetry fractionalization} labeled by
$k \in \Z_{6/{\rm q}}$ 
between 0-form baryon minus lepton U(1)$_{\bf B - L}$ and 1-form electric symmetry is introduced by [arXiv:2411.18160], so we have different versions of 
SM$_{({\rm q},k)}$ with $({\rm q},k)$
dependence.
In this work,
we introduce an integer series of new U(1) of 
baryon {\bf B} 
minus lepton {\bf L}
number like $X_n \equiv n (\mathbf{B}-\mathbf{L}) + (1-\frac{n}{N_c})\tilde{Y}$ with electroweak hypercharge $\tilde{Y}$, $n\ge1$, $N_c=3$, where the charge $q_{X_n} = q_{\tilde{Y}} \mod n$.
We also consider generic compatible non-spin manifolds such as 
$X_n$ requires Spin$^c$ for odd $n$.
We study the Symmetry-Enriched
SM labeled by 
$(G_{[0]},G_{[1]},\rho,[\beta])$.
Here the 0-symmetry is $G_{[0]}=\Spin^c$ for odd $n$ or $\Spin \times \U(1)$ for even $n$;  the 1-symmetry is $G_{[1]}=G_{[1]}^e \times G_{[1]}^m = \Z_{6/{\rm q},[1]}^e \times \U(1)_{[1]}^m$. 
The $\rho: \pi_1(\B G_{[0]}) \to {\rm Aut}(G_{[1]})$ is trivial,
such that 
$[\beta]=([\beta^e],[\beta^m])$
contains the trivial 
electric
$[\beta^e] 
\in [\B G_{[0]}, \B^3 G_{[1]}^e]_{\rho}=0$
twisted homotopy class (thus the zero $\H^3_{\rho}(\B G_{[0]},G_{[1]}^e)$ obstruction class)
and a nontrivial magnetic 
$[\beta^m] 
\in [\B G_{[0]}, \B^3 G_{[1]}^m]_{\rho} = 
\H^4_{\rho}(\B G_{[0]}, \Z)=\Z^2$
obstruction class.
Thus, the symmetry fractionalization
in the electric sector
$k \in  
[\B G_{[0]}, \B^2 G_{[1]}^e]_{\rho}
=
\H^2_{\rho}(\B G_{[0]}, G_{[1]}^e)
= \Z_{6/{\rm q}}$
can be defined.
Due to the mixed anomaly 
$-\frac{1}{2\pi} \tilde{B}_e  \dd B_m$
between the
1-symmetries 
$G_{[1]}^e$ and $G_{[1]}^m$,
the symmetry fractionalization
$\tilde{B}_e \sim \frac{k}{6/{\rm q}}\dd A_{X_n}$ demands a 4d fractional
SPTs
$\frac{1}{2\pi} \frac{k}{6/{\rm q}} 
A_{X_n} \dd B_m$.
In addition,
the spacetime-internal structure imposes the {\bf gauge bundle constraint} 
between $\U(1)_{\tilde{Y}}$ flux,
$\U(1)_{X_n}$ flux, and Stiefel-Whitney classes of spacetime and SM:
$\frac{\dd a_{\tilde{Y}}}{2\pi}=\frac{1}{\text{lcm}(2,n)}\frac{\dd A_{X_n}}{2\pi}-\frac{\text{gcd}(2,n)}{2} w_2(TM) +\frac{1}{{\rm q}}w_2^{({\rm q})}\mod1$. 
Without turning on the background gauge field $B_e$ of the electric 1-symmetry, the magnetic 2-current is $J_m^{(2)}=\star{\rm q}\frac{\dd a_{\tilde{Y}}}{2\pi}$. Substituting the gauge bundle constraint into the coupling $B_m\star J_m^{(2)}$, we obtain a 4d fractional
SPTs
or fractional cobordism invariant $\frac{{\rm q}(1-n)\text{gcd}(2,n)}{2n}\frac{1}{2\pi}{B_m\dd A_{X_n}}$.
Given an SM$_{({\rm q},k)}$ with a choice of $X_{n}$ symmetry probe,
we derive the 
fractional topological response
$\sigma_n({\rm q}, k)=\frac{{\rm q}(1-n)\text{gcd}(2,n)}{2n}+\frac{k{\rm q}}{6}\mod1$;
such that for a fixed $n$ specified by $X_n$,
the $\sigma_n$ can uniquely determine the
SM gauge group ${\rm q}=1,2,3,6$
and fractionalization class $k$, if and only if $n\ge7$ and $n\ne 10,12,15,30$.
Moreover, by choosing a pair of $n=2$ and $3$ together, then
we can further uniquely discern SM$_{({\rm q},k)}$ by measuring both of their $\sigma_2$ and $\sigma_3$. Similarly, pairs of
$\sigma_{n_1}$ and $\sigma_{n_2}$ with $(n_1,n_2)=(2,3)$, $(2,5)$, $(3,4)$, $(3,5)$, $(4,5)$, etc., all such 
pairs can discern SM$_{({\rm q},k)}$.
Our results illuminate the global structure of the SM gauge group via measurable topological responses.

\end{abstract}

%\pacs{}

\maketitle

\begin{center}
  \small Dedicated to Professor Shing-Tung Yau \includegraphics[scale=0.2]{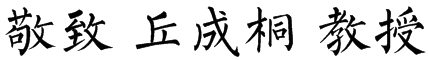} and his 76th Birthday celebration on April 4th, 2025.
\end{center}

%\end{titlepage}

%\onecolumngrid

\tableofcontents

\section{Introduction and Summary}

The Standard Model (SM) of particle physics \cite{Glashow1961trPartialSymmetriesofWeakInteractions, Salam1964ryElectromagneticWeakInteractions, Salam1968, Weinberg1967tqSMAModelofLeptons}, while remarkably successful for describing various subatomic phenomena, harbors a nuanced structure when scrutinized through the lens of global topology \cite{Freed2016}, generalized symmetry \cite{Gaiotto2014kfa1412.5148},
and nonperturbative quantum  regularization \cite{WangWen2018cai1809.11171}.
Recent works in this direction scrutinize the generalized
symmetries and anomalies of the SM and make progress hand in hand together with the development of the deformation class
and cobordism class of the SM (or quantum field theory in general)
\cite{GarciaEtxebarriaMontero2018ajm1808.00009, WangWen2018cai1809.11171, 
WanWang2018bns1812.11967,
NSeiberg-Strings-2019-talk,
Witten2019bou1909.08775,
McNamara2019rupVafa1909.10355,
DavighiGripaiosLohitsiri2019rcd1910.11277, WW2019fxh1910.14668, 
JW2006.16996, JW2008.06499, JW2012.15860, WangWanYou2112.14765, WangWanYou2204.08393, Putrov:2023jqi2302.14862}. 

Although the 
SM, a 4d chiral gauge theory with Yang-Mills spin-1 gauge fields, has
the well-known local Lie algebra, 
\bea \label{eq:SMLieAlgebra}
\cG_{\rm SM} \equiv su(3) \times  su(2) \times u(1)_{\tilde Y},
\eea
we do not yet know which of the four compatible global Lie group structures is nature's choice for the SM
\cite{AharonyASY2013hdaSeiberg1305.0318, Tong2017oea1705.01853}:
\bea \label{eq:SM-Lie-group}
G_{\SM_{\rm q}} \equiv \frac{\SU(3) \times   \SU(2) \times \U(1)_{\tilde Y}}{\Z_{\rm q}},  \quad \text{ with } {\rm q}=1,2,3,6.
\eea
Here ${\tilde Y}$ is a properly quantized electroweak hypercharge for quarks and leptons, see \Table{table:SM}.
The reason is that all four versions in \eqq{eq:SM-Lie-group}
is compatible with the representation (rep) of the 3 families of 15 Weyl fermions of quarks and leptons 
(below written as left-handed 2-component Weyl spinors of the spacetime Spin group)
confirmed by experiments, 
possibly with or without the 16th Weyl fermion (the right-handed neutrino sterile to SM gauge force):
\bea \label{eq:SMrep}
&& \bar{d}_R \oplus {l}_L  \oplus q_L  \oplus \bar{u}_R \oplus   \bar{e}_R  
\oplus
n_{\nu_{j,R}} {\bar{\nu}_{j,R}}\cr
&&\sim 
(\overline{\bf 3},{\bf 1})_{2} \oplus ({\bf 1},{\bf 2})_{-3}  
\oplus
({\bf 3},{\bf 2})_{1} \oplus (\overline{\bf 3},{\bf 1})_{-4} \oplus ({\bf 1},{\bf 1})_{6}  \oplus n_{\nu_{j,R}} {({\bf 1},{\bf 1})_{0}}
\text{ for each family of } su(3) \times su(2) \times u(1)_{\tilde Y}.
\eea
Here $n_{\nu_{j,R}} \in \{0,1\}$ in the presence or absence of the  ${\nu_{j,R}}$ where $j=e,\mu,\tau$ types.

\begin{table}[!h]
\begin{tabular}{|c |  c  | c | c |  c |  c | c  | c|}
%\begin{tabular}{c   c   c  c   c   c  c }
\hline\rule{0pt}{10pt}
& 
$\bar{d}_R = d_L$ & $l_L$ & $q_L$ & $\bar{u}_R = u_L$
& $\bar{e}_R= e_L^+$ & $\bar{\nu}_R= {\nu}_L $ &$\phi_H$\\
\hline\rule{0pt}{10pt}
${\SU(3)}$ & $\overline{\mathbf{3}}$ & $\mathbf{1}$ & ${{\mathbf{3}}}$ & $\overline{\mathbf{3}}$ & $\mathbf{1}$ & $\mathbf{1}$ & $\mathbf{1}$\\
${\SU(2)}$ & $\mathbf{1}$ & $\mathbf{2}$ & $\mathbf{2}$  & $\mathbf{1}$ & $\mathbf{1}$ & $\mathbf{1}$ & $\mathbf{2}$\\
$\U(1)_{Y}$ & 1/3 & $-1/2$ & 1/6 & $-2/3$ & 1 & 0 & ${1}/{2}$\\
$\U(1)_{\tilde Y }$ & 2 & $-3$ & 1 & $-4$ & 6 & 0 & $3$\\
$\U(1)_{\rm{EM}}$ & 1/3 & 0 or $-1$ & 2/3 or $-1/3$ & $-2/3$ & 1 & 0 & 0\\[1mm]
 \hline\rule{0pt}{10pt}
$\U(1)_{{ \mathbf{B}-  \mathbf{L}}}
=\U(1) ^\rF$ & $-1/3$ & $-1$ & $1/3$ & $-1/3$ & 1 & 1 & 0\\
$\U(1)_{{{\bf Q}} - {N_c}{\bf L}}
=\U(1) ^\rF$ & $-1$ & $-3$ & 1 & $-1$ & 3 & 3 & 0\\
$\U(1)_{X} =\U(1) ^\rF$ 
& $-3$ & $-3$ & 1 & 1 & 1 & 5 & $-2$\\
$\Z_{5,X}$ & 2 & 2 & 1 & 1 & 1 & 0 & $-2$\\
$\Z_{4,X}= \Z_4^\rF$ & 1 & 1 & 1 & 1 & 1 & 1 & $-2$\\
$\Z_{8,X}= \Z_8^\rF$ & 5 & 5 & 1 & 1 & 1 & 5 & $-2$\\
$\U(1)_{X_n}$ & $2 - n$ & $-3$ & 1 & $-4 + n$ & $6 - n$ & $n$ & $3 - n$ \\
{\parbox{3.8cm}{ 
\vspace{2pt}
$\Z_{ 2N_f=6, {{\bf B}} + {\bf L}}
=  \Z_6^\rF$ \\
for $N_f=3$;\\
or $\Z_2^\rF, \Z_4^\rF$\\
for $N_f=1,2$\\
(broken from 
$\U(1)_{{{\bf B}} + {\bf L}}$).
\vspace{2pt}}}
& $-1/3$ & $1$ & $1/3$ & $-1/3$ & $-1$ & $-1$ & 0\\
{\parbox{3.8cm}{ 
\vspace{2pt}
$\Z_{ 2N_cN_f=18, {{\bf Q}} + {N_c}{\bf L}}
= \Z_{18}^\rF$ \\
for $N_f=3$;\\
or $\Z_6^\rF, \Z_{12}^\rF$\\
for $N_f=1,2$\\
(broken from 
$\U(1)_{{{\bf Q}} + {N_c}{\bf L}}$).
\vspace{2pt}}}
& $-1$ & $3$ & 1 & $-1$ & $-3$ & $-3$ & 0\\
$\Z_{2}^\rF$ & 1 & 1 & 1 & 1 & 1 & 1 & 0\\
\hline
\end{tabular}
\caption{Follow the convention in \cite{Putrov:2023jqi2302.14862}, we show the representations of quarks and leptons in terms of  
Weyl fermions 
in various internal symmetry groups.
Each fermion is a spin-$\frac{1}{2}$ Weyl spinor 
${\bf 2}_L$ representation {of} the spacetime symmetry group Spin(1,3).
Each fermion is written as a left-handed particle $\psi_L$ or a right-handed anti-particle $\ii \sigma_2 \psi_R^*$.
The groups above the horizontal line are the gauge groups in the SM energy scale.
The groups below the horizontal line are potential global symmetries at the  SM energy scale, but they might be secretly dynamically gauged at higher energies if they are anomaly-free.
Below are some special symmetries, with the color $N_c=3$ in the SM.
The vector $\U(1)_{{{\bf Q}} - {N_c}{\bf L}}$ is the vector $\U(1)_{\mathbf{B}-  \mathbf{L}}$ global symmetry normalized such that the quarks have integer charges.
The 
$ X \equiv 5({ \mathbf{B}-  \mathbf{L}})-\frac{2}{3} {\tilde Y}
  \equiv \frac{5}{N_c}({ \mathbf{Q}-  N_c \mathbf{L}})-\frac{5- N_c}{N_c} {\tilde Y}$ is Wilczek-Zee's chiral symmetry \cite{Wilczek1979hcZee} with the electroweak hypercharge ${\tilde Y}$ being chiral. 
We introduce a new inteer series of $X_n \equiv 
n ({\bf B}-{\bf L}) + (1 - \frac{n}{N_c}) \tilde{Y}
\equiv \frac{n}{N_c}({ \mathbf{Q}-  N_c \mathbf{L}})
 + (1 - \frac{n}{N_c}){\tilde Y}$ labeled by an integer $n \geq 1$.
}
\label{table:SM}
\end{table}

In this work, 
partially inspired by a recent attempt by Hsin-Gomis \cite{Hsin:2024lya2411.18160},
and
partially following some of 
our previous attempts \cite{WangWen2018cai1809.11171, 
WW2019fxh1910.14668, 
JW2006.16996, JW2008.06499, JW2012.15860, WangWanYou2112.14765, WangWanYou2204.08393, Putrov:2023jqi2302.14862},
we shall focus on a  
specific question: {\bf\emph{What is the topological response of 
the 4d SM}} that distinguishes 
four versions of Lie group in \eq{eq:SM-Lie-group}? Here the topological response is in the sense of the response of fractional quantum Hall physics \cite{Stormer:1999zzTsuiGossardRevModPhys, 0707.1889} due to the need of probing the global topology properties of the Lie group structure in \eqq{eq:SM-Lie-group}.

Before attempting to solve this problem of SM, 
we shall sharpen the question further.
We need to first specify the spacetime-internal symmetries, following \cite{Freed2016}, typically written as
\bea
G\equiv ({\frac{{G_{\text{spacetime} }} \times  {{G}_{\text{internal}} } }{{N_{\text{shared}}}}}) \equiv {{G_{\text{spacetime} }} \times_{{N_{\text{shared}}}}  {{G}_{\text{internal}} } }.
\eea
The ${N_{\text{shared}}}$ is the shared common normal subgroup symmetry between ${G_{\text{spacetime} }}$ 
and ${{G}_{\text{internal}} }$, 
e.g. ${N_{\text{shared}}}$ can be the fermion parity symmetry $\Z_2^{\rF}$, which acts on all fermions by $\psi \mapsto - \psi$.
We also include the baryon minus lepton ${ \mathbf{B}-  \mathbf{L}}$ vector symmetry respected by the SM dynamical gauge structure and Yukawa-Higgs terms.

In this work, we discover 
a new continuous $\U(1)_{X_n}$ (see \Sec{sec:Xn}) that is ${ \mathbf{B}-  \mathbf{L}}$ like symmetry,
\bea \label{eq:Xn}
X_n \equiv n({ \mathbf{B}-  \mathbf{L}})+(1-\frac{n}{N_c}) {\tilde Y}
\equiv \frac{n}{N_c}({ \mathbf{Q}-  N_c \mathbf{L}})
 + (1 - \frac{n}{N_c}){\tilde Y},
\eea
where the color $N_c=3$ for the strong force $\SU(N_c)$.
This $X_n$ is a generalization of 
$X_{n=3}= N_c ({ \mathbf{B}-  \mathbf{L}})
={ \mathbf{Q}-  N_c \mathbf{L}}$ symmetry 
and
$X=X_{n=5}
=5({ \mathbf{B}-  \mathbf{L}})-\frac{2}{3} {\tilde Y}
\equiv \frac{5}{3}({ \mathbf{Q}-  N_c \mathbf{L}})
 - \frac{2}{3}{\tilde Y}$ symmetry in Wilczek-Zee's \cite{Wilczek1979hcZee}.
We have decided not to include 
discrete charge conjugation, parity, or time-reversal (C, P, or T) symmetries
as they are already individually broken at the SM energy scale.
The SM is a Lorentz invariant QFT that obeys the spacetime rotation and boost symmetry %($3 + 3$ generators)
${G_{\text{spacetime} }} = \Spin$ group %which is a double cover of the bosonic special orthogonal SO group graded by the fermion parity symmetry $\Z_2^F$.
(Spin(3,1) in Lorentz signature and Spin(4) in Euclidean signature, which is a double cover of SO group via $\Z_2^\rF$ extension).
Gathering the above data, we learn that one physically pertinent version of spacetime-internal symmetries of SM is:
\bea \label{eq:spacetime-internal}
G^{\U(1)_{ X_n}}_{\SM_{\rm q}} &\equiv& {\Spin\times_{\Z_2^{\rF}}\big(\frac{\U(1)_{ X_n }\times G_{\SM_{\rm q}}}{\Z_{n}}\big)}\text{ for odd }n,\nn\\
G^{\U(1)_{ X_n}}_{\SM_{\rm q}} &\equiv& {\Spin\times\big(\frac{\U(1)_{ X_n }\times G_{\SM_{\rm q}}}{\Z_{n}}\big)}\text{ for even }n.  
\eea
Here, there is a shared common subgroup $\Z_n$ between
$\U(1)_{X_n}$ and $\U(1)_{ \tilde Y }$, 
because the charges of $\U(1)_{X_n}$ and $\U(1)_{ \tilde Y }$ are equal modulo $n$ for all particles in \Table{table:SM}. 
Namely the
charges 
\bea
q_{X_n} = q_{\tilde{Y}}
\mod n 
\eea
are identical modulo $n$.

For odd $n$, there is a shared common subgroup $\Z_2^{\rF}$ between $\Spin$ and $\U(1)_{X_n}$; because the charge of $\mathbf{B}-\mathbf{L}-\frac{\tilde{Y}}{3} \in \Z$ is an integer for all particles in \Table{table:SM}, hence the charges of $\U(1)_{X_n}$ and $\U(1)_{X_3=\mathbf{Q}-3\mathbf{L}}$ differ by
$(n-3)(\mathbf{B}-\mathbf{L}-\frac{\tilde{Y}}{3}) \in 2 \Z$, 
thus always have the same parity (even or odd) for all particles in \Table{table:SM} for odd $n$. Since the charge of $\U(1)_{X_3=\mathbf{Q}-3\mathbf{L}}$ is odd for all fermions in \Table{table:SM} and even for the Higgs boson in \Table{table:SM}, so is the charge of $\U(1)_{X_n}$ for odd $n$. 

For even $n$, there is no shared common subgroup $\Z_2^{\rF}$ between $\Spin$ and $\U(1)_{X_n}$; because the charge of $\mathbf{B}-\mathbf{L}-\frac{\tilde{Y}}{3} \in \Z$ is an integer for all particles in \Table{table:SM}, hence the charges of $\U(1)_{X_n}$ and $\U(1)_{\tilde{Y}}$ differ by
$n(\mathbf{B}-\mathbf{L}-\frac{\tilde{Y}}{3}) \in 2 \Z$, 
thus always have the same parity (even or odd) for all particles in \Table{table:SM} for even $n$. Since the charge of $\U(1)_{\tilde{Y}}$ is not necessarily odd for all fermions in \Table{table:SM} and not even for the Higgs boson in \Table{table:SM}, so is the charge of $\U(1)_{X_n}$ for even $n$.  

Therefore, this spacetime-internal symmetry \eqref{eq:spacetime-internal} is faithful.
The symmetry \eqq{eq:spacetime-internal} 
is still compatible with the rep \eq{eq:SMrep}
of the 3 families of 15 Weyl fermions, 
possibly with or without the 16th Weyl fermion.

In a gauge theory,
the {internal symmetry} also includes \emph{higher $n$-symmetries} as generalized global symmetries \cite{Gaiotto2014kfa1412.5148}
that act on $n$d charged objects (1d lines, 2d surfaces, etc.)
by coupling the $n$d objects to $n+1$d background gauge fields.
We denote the higher $n$-symmetry group $G$ as $G_{[n]}$.
The $n+1$d gauge field lives in a higher classifying space $\B^{n+1} G$ (namely, the Eilenberg-MacLane space $K(G,n+1)$ when $G$ is discrete)
of the group $G_{[n]}$.  %
Once $G_{\text{SM}_{\rm q}}$ is gauged, 
\emph{kinematically}, we may obtain extra 1-form electric and magnetic symmetries, $G_{[1]}^e$ and $G_{[1]}^m$,
which are determined from the center of gauge group $Z(G_{\SM_{\rm q}})$,
{and the Pontryagin dual of the homotopy group $\pi_1(G_{\SM_{\rm q}})^{\vee}=\Hom(\pi_1(G_{\SM_{\rm q}}),\U(1))=\Hom( \Z,\U(1))=\U(1)$.}
Moreover the 0d SM gauge charge fermionic particle \eqq{eq:SMrep} explicitly breaks the $G_{[1]}^e$ from $Z(G_{\SM_{\rm q}})$ to a subgroup 
\cite{Wan2019sooWWZHAHSII1912.13504, AnberPoppitz2110.02981, WangYou2111.10369GEQC}
that acts trivially on all particle representations:
\bea \label{eq:1-symmetry}
\begin{tabular}{|c | c |c| c| c |}
\hline
		 & $Z(G_{\SM_{\rm q}})$ &  $\pi_1(G_{\SM_{\rm q}})^{\vee}$ &
		1-form $e$ sym $G_{[1]}^e$  &
		1-form $m$ sym $G_{[1]}^m$ 
		\\ 
		\hline
		 $G_{\SM_{\rm q}} \equiv \frac{{\SU(3)} \times {\SU(2)} \times \U(1)_{\tilde{Y}}}{\Z_{\rm q}}$ & 
		 $\Z_{6/{\rm q}} \times \U(1)$ & $\U(1)$ &    $\Z_{6/{\rm q},[1]}^e$ & ${\U(1)}_{[1]}^m$  \\
		\hline
\end{tabular}, \quad 
{\rm q}=1,2,3,6.
\eea
Thus, 
starting from the ungauged spacetime-internal symmetry
\eq{eq:spacetime-internal},
by dynamically gauging $G_{\SM_{\rm q}}$,
we have to remove $G_{\SM_{\rm q}}$ and include $G_{[1]}^e \times G_{[1]}^m$
 into \eq{eq:spacetime-internal}, so the gauged SM
has the following spacetime-internal symmetry:
\bea 
\label{eq:spacetime-internal-gauged}
G'^{\U(1)_{ X_n }}_{\SM_{\rm q}} &\equiv& \Spin \times_{\Z_2^F} \frac{\U(1)_{ X_n }}{\Z_{n}} \times \Z_{6/{\rm q},[1]}^e \times \U(1)_{[1]}^m\text{ for odd }n,\nn\\
G'^{\U(1)_{ X_n }}_{\SM_{\rm q}} &\equiv& \Spin \times \frac{\U(1)_{ X_n }}{\Z_{n}} \times \Z_{6/{\rm q},[1]}^e \times \U(1)_{[1]}^m\text{ for even }n. 
\eea
N\"aively,
we obtain
$G_{[1]}=G_{[1]}^e \times G_{[1]}^m = \Z_{6/{\rm q},[1]}^e \times \U(1)_{[1]}^m$ 
\cite{Wan2019sooWWZHAHSII1912.13504, AnberPoppitz2110.02981, WangYou2111.10369GEQC, WangWanYou2112.14765, WangWanYou2204.08393}.
However, there is a possibility of symmetry-fractionalization
between 0-symmetry and 1-symmetry in the SM
recently pointed out by Ref.~\cite{Hsin:2024lya2411.18160}.

Follow the general procedure of the symmetry fractionalization (e.g. 
\cite{Barkeshli:2014cna1410.4540,
Chen:2016fxq1606.07569, 1803.09336}),
we will compute $(G_{[0]},G_{[1]},\rho ,[\beta])$
explicitly, 
then compute the SM's symmetry fractionalization 
$k =(k^e, k^m)$ in electric and magnetic sectors,
detailed in \Sec{sec:symmetry-fractionalization} and \App{sec:SETSM}.
The 0-symmetry and 1-symmetry of the SM are given by 
\bea
(G_{[0]}, G_{[1]})
=(G_{[0]}, G_{[1]}^e \times G_{[1]}^m  )
=
(\left\{\begin{array}{l}
 \Spin \times_{\Z_2^F} \frac{\U(1)_{ X_n }}{\Z_{n}}  = \Spin^c, \quad n \in \Z_{\rm odd} \\
\Spin \times \frac{\U(1)_{ X_n }}{\Z_{n}} = \Spin \times \U(1), \quad n \in \Z_{\rm even}
\end{array} \right.,
\quad
\Z_{6/{\rm q},[1]}^e \times \U(1)_{[1]}^m) 
\eea 
 The symmetry twist is a group homomorphism $\rho$
 (from the first homotopy group of classifying space $\B G_{[0]}$ to the automorphism group of $G_{[1]}$)
which is trivial for the SM:
\bea
\rho: \pi_1(\B G_{[0]}) \to {\rm Aut}(G_{[1]})
\Rightarrow \rho_{\rm SM}=0, 
\eea
which shows how the 0-form symmetry's codimension-1 topological defect acts on the 1-form symmetry's charged line.
The obstruction of the symmetry-fractionalization 
is determined by the homotopy class
\bea
[\beta] = ([\beta^e], [\beta^m])  \in [\B G_{[0]}, \B^3 G_{[1]}]_{\rho} = ([\B G_{[0]}, \B^3 G_{[1]}^e]_{\rho}, [\B G_{[0]}, \B^3 G_{[1]}^m]_{\rho}) = (0, \Z^2).
\eea
%\jw{continue from here?}
If $[\beta^m]$ is nontrivial, the magnetic symmetry fractionalization can not be defined.
If $[\beta^m]$ is trivial, namely there is no obstruction, the magnetic symmetry fractionalization class $k^m\in [\B G_{[0]}, \B^2 G_{[1]}^m]_{\rho}=0$, thus there is no magnetic symmetry fractionalization.
Since $[\beta^e]$ is trivial, the electric symmetry fractionalization class 
\bea 
k^e\in [\B G_{[0]}, \B^2 G_{[1]}^e]_{\rho}=\Z_{6/{\rm q}}.
\eea 
We simply denote $k \equiv (k^e, k^m)$ as $k \equiv k^e$ since $k^m=0$. 
See \Sec{sec:symmetry-fractionalization} for a summary 
of the SM symmetry fractionalization and \Fig{fig:flow} for the flow chart of the logical steps.
See \App{sec:SETSM} for the mathematics of symmetry fractionalization.

Thus, the symmetry fractionalization $k=k_e \in
\Z_{6/{\rm q}}$ requires the following constraint 
between the background fields of electric 1-form symmetry
$G_{[1]}^e  = \Z_{6/{\rm q},[1]}^e$ and the 0-form symmetry $G_{[0]}$:
\bea
\tilde{B}_e\sim \frac{k}{6/{\rm q}}\dd A_{X_n}.
\eea
Here $\tilde{B}_e=\frac{2\pi}{6/{\rm q}}B_e$, with $B_e\in \H^2(\B^2\Z_{\frac{6}{{\rm q}}},\Z_{\frac{6}{{\rm q}}})=\Z_{\frac{6}{{\rm q}}}$ is the background gauge field of the electric 1-form $\Z_{6/{\rm q},[1]}^e$ symmetry of the SM, while $A_{X_n}\in\Omega^1(\B(\frac{\U(1)_{ X_n }}{\Z_{\text{lcm}(2,n)}}),\R)$ is the background gauge field of the $\frac{\U(1)_{X_n}}{\Z_{\text{lcm}(2,n)}}$ symmetry, and $\sim$ means that the fields on both sides are equal when they are pulled back to the spacetime manifold $M^4$, see \Sec{sec:SF-odd} and \Sec{sec:SF-even}.

 There is a mixed anomaly \cite{WangWanYou2112.14765, WangWanYou2204.08393}
  between the electric 1-form symmetry and the magnetic  1-form symmetry 
 $G_{[1]}^e \times G_{[1]}^m = \Z_{6/{\rm q},[1]}^e \times \U(1)_{[1]}^m$ of the Standard Model,
\bea \label{eq:5dBdB}
\int_{M^5}-\frac{1}{2\pi}\tilde{B}_e\dd B_m
= \int_{M^5}-\frac{1}{6/{\rm q}}B_e \dd B_m ,
\eea
 for ${\rm q}=1,2,3,6$, where $B_m\in \Omega^2(\B^2\U(1),\R)$ is the background gauge field of the magnetic 1-form $\U(1)_{[1]}^m$ symmetry of the SM, and $M^5$ is a 5-manifold with the boundary $\partial M^5=M^4$, see \Sec{sec:logic}.

 Due to the mixed anomaly 
$-\frac{1}{2\pi} \tilde{B}_e  \dd B_m$,
the symmetry fractionalization
$\tilde{B}_e \sim \frac{k}{6/{\rm q}}\dd A_{X_n}$ demands a 4d fractional
SPTs
\bea \label{eq:4dAdB}
\int_{M^4} \frac{1}{2\pi} \frac{k}{6/{\rm q}} 
A_{X_n} \dd B_m ,
\eea thanks to the anomaly inflow from the 5d \eqq{eq:5dBdB} 
to the 4d \eq{eq:4dAdB}.
Thus, we shall focus on the topological response 
in the form of \eqq{eq:4dAdB}
\bea 
  \int_{M^4}\frac{\sigma_n}{2\pi}B_m\dd A_{X_n}
\eea
and the ${\sigma_n}({\rm q}, k)$ with $({\rm q}, k)$-dependence 
of the SM gauge group, where $M^4$ is a possibly 
  \emph{non-spin} 4-manifold in our case (that allow 
  the second Stiefel-Whitney class $w_2(TM) \neq 0$
  thus more general than the spin structure required in \cite{Hsin:2024lya2411.18160} restricted to $w_2(TM) = 0$). We derive the fractional part of $\sigma_n$, which consists of two parts: 
  \begin{enumerate}
  \item
  One contribution $\frac{{\rm q}(1-n)\text{gcd}(2,n)}{2n}$
  of the SM's $\sigma_n$ response
  is obtained from the fractional part of $\star J_m^{(2)}$ due to the coupling $B_m\star J_m^{(2)}$ (see footnote \ref{footnote-F}) where $J_m^{(2)}$ is the 2-form current of the magnetic 1-form symmetry 
  $\U(1)_{[1]}^m$  without turning on the background gauge field $B_e$ of the electric 1-form symmetry $\Z_{6/{\rm q},[1]}^e$ by using the gauge bundle constraint (see \Sec{sec:GBC-odd} and \Sec{sec:GBC-even}):
  $$
  \frac{\dd a_{\tilde{Y}}}{2\pi}=\frac{1}{\text{lcm}(2,n)}\frac{\dd A_{X_n}}{2\pi}-\frac{\text{gcd}(2,n)}{2} w_2(TM) +\frac{1}{{\rm q}}w_2^{({\rm q})}\mod 1,
  $$
and the gauge bundle constraint of the gauged SM for odd $n$
\eqref{eq:spacetime-internal-gauged}: 
$$
w_2(TM)=\frac{\dd A_{X_n}}{2\pi}\mod 2.
$$
See \App{app:notation} for the notations used here.
\item Another contribution $\frac{k{\rm q}}{6}$ 
 of the SM's $\sigma_n$ response
is already obtained earlier from the symmetry fractionalization of the 0-form symmetry 
$G_{[0]}$
%{$\mathrm{Spin}\times_{\mathbb{Z}_2^{\mathrm{F}}}\frac{\mathrm{U}(1)_{X_n}}{\mathbb{Z}_n}$ for odd $n$ or $\mathrm{Spin}\times\frac{\mathrm{U}(1)_{X_n}}{\mathbb{Z}_n}$ for even $n$ }
and electric 1-form symmetry 
$G_{[1]}^e  = \Z_{6/{\rm q},[1]}^e$ of the gauged SM with the fractionalization class $k\in\mathbb{Z}_{6/{\rm q}}$, obtained by canceling the mixed anomaly of \eq{eq:5dBdB} (see \Sec{sec:SF-odd} and \Sec{sec:SF-even}). 
\end{enumerate}
Our topological response result of the SM is the combination of the two contributions
\bea
\sigma_n={\sigma_n}({\rm q}, k) = \frac{{\rm q}(1-n)\text{gcd}(2,n)}{2n}+\frac{k{\rm q}}{6}\mod 1.
\eea
Note that this $\sigma_n$ is \emph{independent} of 
the family number $N_f$ (typically $N_f=3$) and the 
sterile neutrino ${\nu_R}$ number 
$n_{\nu_{R}} \equiv \sum_{j=e,\mu,\tau,\dots} n_{\nu_{j,R}}$ 
{that can be equal, smaller, or larger than 3}.

For a fixed $n$ specified by the $X_n$ global symmetry,
the $\sigma_n({\rm q}, k)$ can uniquely determine the
SM gauge group for ${\rm q}=1,2,3,6$
and its fractionalization class $k$, if and only if 
$$
\text{$n\ge7$ and $n\ne 10,12,15,30$.}
$$
Moreover, by choosing a pair of $n=2$ and $3$ together, then
we can further uniquely discern SM$_{({\rm q},k)}$ by measuring both of their $\sigma_2$ and $\sigma_3$. Similarly, pairs of
$\sigma_{n_1}$ and $\sigma_{n_2}$ with 
$$
\text{$(n_1,n_2)=(2,3)$, $(2,5)$, $(3,4)$, $(3,5)$, $(4,5), \dots$, etc.,} 
$$ 
all such 
pairs can discern SM$_{({\rm q},k)}$, see \Table{table:response}.
Thus, in comparison with \cite{Hsin:2024lya2411.18160},
Hsin-Gomis shows only the special case of our $n=3$ that cannot discern completely the four versions of SM$_{({\rm q},k)}$ with ${\rm q}=1,2,3,6$.
For example, at $n=3$ in \Table{table:response},
$\sigma_3{({\rm 1},2)}=\sigma_3{({\rm 2},2)}=\sigma_3(3,0)=\sigma_3(6,0)=0$, so the topological response cannot distinguish SM$_{({\rm 1},2)}$, SM$_{({\rm 2},2)}$, SM$_{(3,0)}$, and SM$_{(6,0)}$.

In summary, our introduction of $\U(1)_{X_n}$ symmetry
in \eq{eq:Xn} for generic $n$ in \Table{table:response}
extends the results of \cite{Hsin:2024lya2411.18160}, 
enabling a complete distinction of the SM variants
SM$_{({\rm q},k)}$ completely.\\

The plan of this article is organized as follows:

In \Sec{sec:Xn}, we define a new global symmetry $\U(1)_{X_n}$ of the Standard Model.

In \Sec{sec:symmetry-fractionalization}, we study the symmetry fractionalization of the Standard Model:
0- and 1-symmetry $(G_{[0]},
G_{[1]})$, homomorphism 
$\rho$, obstruction $[\beta]$,
and fractionalization class
$k=(k^e, k^m)$.

In \Sec{sec:generic}, we study the generic topological response theory using gauge bundle constraint and symmetry fractionalization.

In \Sec{sec:conclusion}, we discuss for which values of $n$, the topological response $\sigma_n$ can distinguish the four versions of SM$_{({\rm q},k)}$ with ${\rm q}=1,2,3,6$ .

In \App{app:notation}, we introduce the systematic notations and definitions used in this article.

In \App{app:classifying} and \App{app:higher}, we introduce the classifying space and higher classifying space.

In \App{app:cohomology}, we compute the second and third integral cohomology of $\B\Spin^c$ and $\B(\Spin \times \U(1))$.

In \App{sec:SETSM},
we explore the 
Symmetry-Enriched Standard Model SM$_{(\rm {\rm q},k)}$:
$(G_{[0]},\; 
G_{[1]},\; \rho,\; [\beta])$ with $\rho: \pi_1(\B G_{[0]}) \to {\rm Aut}(G_{[1]})$ 
and $[\beta] \in  
[\B G_{[0]}, \B^3 G_{[1]}]_{\rho}$,
and symmetry fractionalization 
$k \in  
[\B G_{[0]}, \B^2 G_{[1]}]_{\rho}$ when $[\beta]=0$.

In \App{sec:cobordism}, we compute the bordism groups $\Omega_d^{\Spin^c\times\B\Z_{6/{\rm q}}\times\B\U(1)}$ and the TP groups $\TP_d(\Spin^c\times\B\Z_{6/{\rm q}}\times\B\U(1))$ for ${\rm q}=2,3$ and $d\le5$.
We also list down cobordism invariants also known as 
the invertible field theories (iFT)  
as symmetry-protected topological states' (SPTs) invariants
\cite{1405.7689, Kapustin1406.7329}.
Some of these iFTs are indeed 
\emph{metric-independent} thus those iFTs can be invertible topological field theories (iTFT) in the {\bf mathematical} sense of topology.
The topology in the SPTs is 
in the {\bf physical} sense of being \emph{gapped}
in the energy spectrum.

\section{A New Modified Baryon minus Lepton Symmetry $X_n \equiv 
n ({\bf B}-{\bf L}) + (1 - \frac{n}{N_c}) \tilde{Y} \equiv \frac{n}{N_c}({ \mathbf{Q}-  N_c \mathbf{L}})
 + (1 - \frac{n}{N_c}){\tilde Y}$}\label{sec:Xn}

While the $\mathbf{B}-  \mathbf{L}$ is a well-known global symmetry at the SM scale and the electroweak hypercharge $\tilde{Y}$ is a dynamical gauge charge,
we can still linearly combine the global $\mathbf{B}-  \mathbf{L}$
and the gauge $\tilde{Y}$ appropriately as a new
global symmetry if we restrict the $\tilde{Y}$ gauge transformation as a global symmetry transformation (independent of the spacetime coordinates).
This logic is used in  Wilczek-Zee's 
$X=5({ \mathbf{B}-  \mathbf{L}})-\frac{2}{3} {\tilde Y}$ symmetry \cite{Wilczek1979hcZee}.

In this section, we define a new global symmetry $\U(1)_{X_n}$ such that 
\bea
X_n \equiv n(\mathbf{B}-  \mathbf{L}) + \lambda \tilde{Y}, 
\eea 
with $X_n$ charges
for quarks and leptons fermions, and Higgs boson
$(\bar{d}_R, 
{l}_L, 
q_L,
\bar{u}_R, 
\bar{e}_R,  
{\bar{\nu}_{j,R}},
\phi_H)$
as
\bea\label{eq:charges}
q_{X_n}=(-n/3+2 \lambda, -n-3\lambda, n/3 + \lambda, -n/3 - 4 \lambda, n+6\lambda, n, 3 \lambda).
\eea
Here 
the charges and 
representations of quarks and leptons in terms of  
Weyl fermions, and Higgs boson, 
in various gauge groups and internal symmetry groups are given in \Table{table:SM}.

We require that all of the above charges are integers so that quarks and leptons fermions, and Higgs boson carry integer charges under $\U(1)_{X_n}$, and at least one of them is $\pm 1$, so it means that before promoting the $G_{\SM_{\rm q}}$ to dynamical gauge group, it is a faithful symmetry. We find a solution,\footnote{We derive the general solutions as follows.
\begin{enumerate}
    \item Since the $X_n$ charge of $q_L$ is $n/3 + \lambda\in\Z$, let $n/3 + \lambda \equiv p\in\Z$. The $X_n$ charges of the quarks and leptons fermions, and Higgs boson, $(\bar{d}_R, 
{l}_L, 
q_L,
\bar{u}_R, 
\bar{e}_R,  
{\bar{\nu}_{j,R}},
\phi_H)$, are 
    $$q_{X_n}=(-n + 2 p,\; -3 p,\; p,\; n - 4 p,\; -n + 6 p,\; n,\; -n + 3 p).$$
    \item Since at least one of the $X_n$ charges of the quarks and leptons fermions, and Higgs boson, $(\bar{d}_R, 
{l}_L, 
q_L,
\bar{u}_R, 
\bar{e}_R,  
{\bar{\nu}_{j,R}},
\phi_H)$, is $\pm 1$, then 
$$
\text{$p=\pm 1$ or $n=\pm1$ or $n-2p=\pm1$ or $n-3p=\pm1$ or $n-4p=\pm1$ or $n-6p=\pm1$}.
$$
\end{enumerate}
We find that the special solution $p=1$ has the advantage that there is a shared common subgroup $\Z_n$ between
$\U(1)_{X_n}$ and $\U(1)_{ \tilde Y }$ because the charge of $\mathbf{B}-\mathbf{L}-\frac{\tilde{Y}}{3}$ is an integer for all particles in \Table{table:SM}, hence the charges of $\U(1)_{X_n}$ and $\U(1)_{ \tilde Y }$ are equal modulo $n$ for all particles in \Table{table:SM}. } which is 
$$\lambda=-n/3+1.$$

So the $X_n$ charges \eqref{eq:charges} are
\bea
q_{X_n}=(2 - n,\; -3, \; 1, \; -4 + n, \; 6 - n, \; n, \; 3 - n).
\eea

There is a shared common subgroup $\Z_n$ between
$\U(1)_{X_n}$ and $\U(1)_{ \tilde Y }$, 
because the charges of $\U(1)_{X_n}$ and $\U(1)_{ \tilde Y }$ are equal modulo $n$ for all particles in \Table{table:SM}. 
Namely the
charges 
\bea
q_{X_n} = q_{\tilde{Y}}
\mod n 
\eea
are identical modulo $n$.
To derive that, 
because the charge of 
\bea
\mathbf{B}-\mathbf{L}-\frac{\tilde{Y}}{N_c} \in \Z
\eea
is an integer for all particles in \Table{table:SM}
(with the color $N_c=3$),  
the combination of that $\mathbf{B}-\mathbf{L}-\frac{\tilde{Y}}{N_c}$
and an integer charge ${\tilde{Y}} \in \Z$
gives rise to a new integer charge $X_n$ symmetry,
\bea
X_n &\equiv& 
n ({\bf B}-{\bf L}) + (1 - \frac{n}{N_c}) \tilde{Y} =
n (\mathbf{B}-\mathbf{L}-\frac{\tilde{Y}}{N_c}) + {\tilde{Y}} \in \Z\cr
&\equiv& 
\frac{n}{N_c}({ \mathbf{Q}-  N_c \mathbf{L}})+ (1 - \frac{n}{N_c}) \tilde{Y}.
\eea

In particular, 
\begin{itemize}

\item 
When $n=3$, it reduces to the $N_c$ times of 
the baryon minus lepton 
symmetry
$$
X_{3}= N_c ({ \mathbf{B}-  \mathbf{L}})
={ \mathbf{Q}-  N_c \mathbf{L}}.  
$$

\item 
When $n=5$, it reduces to 
Wilczek-Zee's $X$ chiral symmetry 
$$ X_5 = X \equiv 5({ \mathbf{B}-  \mathbf{L}})-\frac{2}{3} {\tilde Y}
  \equiv \frac{5}{N_c}({ \mathbf{Q}-  N_c \mathbf{L}}) 
  + (1 - \frac{5}{N_c}) {\tilde Y}.$$  
  
Note that the $X_5$ has two advantages: First, being
consistent with the $su(5)$ grand unification \cite{Georgi1974syUnityofAllElementaryParticleForces}
to have a U(5) group structure, so the $su(5) \times u(1)_{X_5}$ charges for SM fermion multiplet
are $\overline{\bf 5}_{-3} \oplus {\bf 10}_1  \oplus {\bf 1}_5$. Second, the $\Z_{4,X_5}$ charge
as $X_5 \mod 4 = 1 \mod 4$ for SM fermion multiplet,
thus $\Z_{4,X_5} = Z(\Spin(10))$ sits at the center of the Spin(10) gauge group, compatible with the $so(10)$
grand unification \cite{Fritzsch1974nnMinkowskiUnifiedInteractionsofLeptonsandHadrons}. 

\end{itemize}

\section{Symmetry Fractionalization of the Standard Model:\\
0- and 1-symmetry $(G_{[0]},
G_{[1]})$, homomorphism 
$\rho$, obstruction $[\beta]$,
and fractionalization class
$k=(k^e, k^m)$}
\label{sec:symmetry-fractionalization}

\begin{figure}[!h]
\includegraphics[scale=0.5]{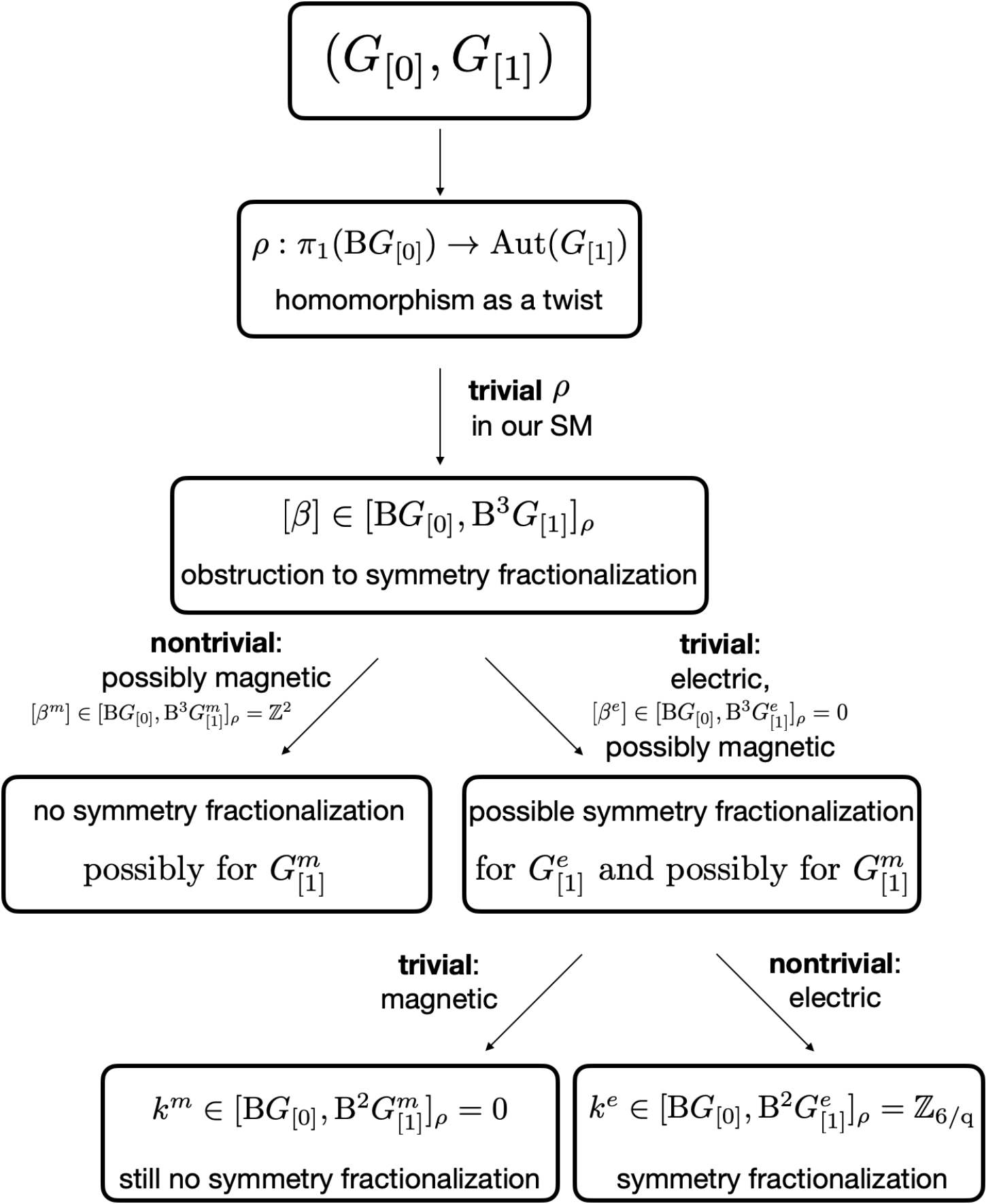}
\caption{The flow chart of the Standard Model (SM)'s symmetry fractionalization $k =(k^e, k^m)$ in electric and magnetic sectors. Given the 0-symmetry and 1-symmetry of the SM, $(G_{[0]}, G_{[1]})
=(\Spin^c \text{ or } \Spin \times \U(1),
\Z_{6/{\rm q},[1]}^e \times \U(1)_{[1]}^m)$ in \eq{eq:spacetime-internal-gauged}, step by step,  we determine (1) the homomorphism
$\rho$ known as a symmetry twist. 
It turns out that our SM \eqq{eq:spacetime-internal-gauged} here has a trivial $\rho=0$. (2) Obstruction to symmetry fractionalization $[\beta] \in [\B G_{[0]}, \B^3 G_{[1]}]_{\rho}$ in the SM 
such that the electric's obstruction $[\beta^e]\in [\B G_{[0]}, \B^3 G_{[1]}^e]_{\rho}=0$ is 
always trivial, while
magnetic obstruction $[\beta^m]\in [\B G_{[0]}, \B^3 G_{[1]}^m]_{\rho}=\Z^2$ can be nontrivial.
(3) Because $[\beta^e]=0$, then the electric symmetry fractionalization $k^e\in[{\rm B}G_{[0]},{\rm B}^2G_{[1]}^e]_{\rho}=\mathbb{Z}_{6/{\rm q}}$ can be defined. Either $[\beta^m] \neq 0$, or $[\beta^m]=0$
but still $k^m\in[{\rm B}G_{[0]},{\rm B}^2G_{[1]}^m]_{\rho}=0$;
for both scenarios, 
there is no magnetic symmetry fractionalization
$k^m=0$. Since we only have 
$k =(k^e, k^m)=(\mathbb{Z}_{6/{\rm q}},0)$,
we may just denote the $k=k^e$ as the nontrivial electric one.
See \App{sec:SETSM}.}
\label{fig:flow}
\end{figure}

Follow the general procedure of the symmetry fractionalization (e.g. 
\cite{Barkeshli:2014cna1410.4540,
Chen:2016fxq1606.07569, 1803.09336}),
we will compute 
\bea
(G_{[0]},G_{[1]},\rho ,[\beta])
\eea
explicitly, 
then compute the SM's symmetry fractionalization class 
$k =(k^e, k^m)$ in electric and magnetic sectors,
see \Fig{fig:flow} for the flow chart,
and see \App{sec:SETSM} for the detailed calculations.
Given the 0-symmetry and 1-symmetry of the SM, 
\bea
(G_{[0]}, G_{[1]})
=(\Spin^c \text{ or } \Spin \times \U(1),\quad
\Z_{6/{\rm q},[1]}^e \times \U(1)_{[1]}^m) 
\eea 
in \eq{eq:spacetime-internal-gauged}, step by step,  we determine: 
\begin{enumerate}
    
\item The homomorphism
\bea
\rho: \pi_1(\B G_{[0]}) \to {\rm Aut}(G_{[1]}). 
\eea  
known as a symmetry twist such that the 0-form symmetry's codimension-1 topological defect acts on the 1-form symmetry's charged line as $\rho_{g_{[0]}}(g_{[1]})$ as in \cite{Barkeshli:2014cna1410.4540}. 
It turns out that our SM \eqq{eq:spacetime-internal-gauged} here has a trivial $\rho=0$. 

\item  Obstruction to symmetry fractionalization 
is given by $[\beta]$, the twisted homotopy class
\bea
[\beta] \in [\B G_{[0]}, \B^3 G_{[1]}]_{\rho}
\eea
in the SM 
such that the electric's obstruction 
\bea
[\beta^e]\in [\B G_{[0]}, \B^3 G_{[1]}^e]_{\rho}=0
\eea
is always trivial, while
magnetic obstruction 
\bea
[\beta^m]\in [\B G_{[0]}, \B^3 G_{[1]}^m]_{\rho}=\Z^2
\eea
can be nontrivial.

\item  Because $[\beta^e]=0$, then the electric symmetry fractionalization 
\bea
k^e\in[{\rm B}G_{[0]},{\rm B}^2G_{[1]}^e]_{\rho}=\mathbb{Z}_{6/{\rm q}}
\eea 
can be defined. Either $[\beta^m] \neq 0$, or $[\beta^m]=0$
but still 
\bea
k^m\in[{\rm B}G_{[0]},{\rm B}^2G_{[1]}^m]_{\rho}=0;
\eea 
for both scenarios, 
there is no magnetic symmetry fractionalization
$k^m=0$. Since we only have 
\bea
k =(k^e, k^m)=(\mathbb{Z}_{6/{\rm q}},0),
\eea
we may just denote the $k=k^e$ quoting only the nontrivial electric one.

\end{enumerate}

%\newpage

\section{Generic Topological Response Theory}\label{sec:generic}

\subsection{Logic and arguments}\label{sec:logic}

Let us give an overview of the logic and arguments required to derive a 
generic topological response theory for the SM.
In \Sec{sec:GBC-odd} and \Sec{sec:GBC-even}, we derive the gauge bundle constraint
\bea\label{eq:constraint-X}
\frac{\dd a_{\tilde{Y}}}{2\pi}=\frac{1}{\text{lcm}(2,n)}\frac{\dd A_{X_n}}{2\pi}-\frac{\text{gcd}(2,n)}{2} w_2(TM) +\frac{1}{{\rm q}}w_2^{({\rm q})}\mod1
\eea
of the ungauged Standard Model \eqref{eq:spacetime-internal}. Here, $a_{\tilde{Y}}$ is the dynamical gauge field of the $\U(1)_{\tilde{Y}}$ symmetry, $A_{X_n}$ is the background gauge field of the $\frac{\U(1)_{X_n}}{\Z_{\text{lcm}(2,n)}}$ symmetry, $w_2(TM)$ is the second Stiefel-Whitney class of the tangent bundle which is the obstruction class to lifting a $\frac{\Spin}{\Z_2^{\rF}}$ bundle to a $\Spin$ bundle, and $w_2^{({\rm q})}$\footnote{The number ${\rm q}$ in the superscript indicates that this cohomology class is a mod ${\rm q}$ class. Other similar notations have the same meaning.} is the obstruction class to lifting a $\frac{\SU(3)\times\SU(2)}{\Z_{\rm q}}$ bundle to a $\SU(3)\times\SU(2)$ bundle.

The gauge bundle constraint of the gauged Standard Model for odd $n$
\eqref{eq:spacetime-internal-gauged} is 
\bea\label{eq:constraint-X-gauged}
w_2(TM)=\frac{\dd A_{X_n}}{2\pi}\mod 2.
\eea

By \cite{Gaiotto2014kfa1412.5148}, when the background gauge field $B_e$ for the electric 1-form symmetry $\Z_{6/{\rm q},[1]}^e$ of the Standard Model is not turned on, the magnetic 1-form symmetry $\U(1)_{[1]}^m$ is generated by a 2-form current 
\bea\label{eq:current-X}
J_m^{(2)}= {\rm q}\star \frac{\dd a_{\tilde{Y}}}{2\pi}.
\eea
Let $B_m$ be the background gauge field for the magnetic 1-form symmetry $\U(1)_{[1]}^m$. 
We consider the topological response\footnote{The term $\int_{M^4} \frac{1}{2\pi} B_m\dd A_{X_n}=\int_{M^4}\frac{1}{2\pi}A_{X_n}\dd B_m$ appears as an SPTs or cobordism invariant (free $\Z$ class) in \Table{table:U1-q=3-TP} and \Table{table:U1-q=2-TP}.
While \eq{eq:BmdA} with ${\sigma_n} \mod 1$ is a fraction between 0 and 1, thus
\eq{eq:BmdA} is a fractional SPTs or fractional cobordism invariant.}
\bea\label{eq:BmdA}
\int_{M^4} \frac{\sigma_n}{2\pi} B_m\dd A_{X_n}
\eea
of the Standard Model, this term comes from the Standard Model Lagrangian in two ways. One way is substituting the gauge bundle constraints \eqref{eq:constraint-X} and \eqref{eq:constraint-X-gauged} into the coupling term $B_m\star J_m^{(2)}$\footnote{Let $F$ be the fractional part of $\star J_m^{(2)}$. The difference between the fractional part of $B_m\star J_m^{(2)}$ and $B_mF$ is $B_m(\star J_m^{(2)}-F)$. By \eqref{eq:constraint-X}, \eqref{eq:constraint-X-gauged}, and \eqref{eq:current-X}, $\star J_m^{(2)}-F$ is an integer-valued exact form, hence $\dd(\star J_m^{(2)}-F)=0$. So $\int_{M^4}B_m(\star J_m^{(2)}-F)=\int_{M^5}\dd B_m(\star J_m^{(2)}-F)\in2\pi\Z$ does not affect the SM action because $\exp(\ii \int_{M^4}B_m(\star J_m^{(2)}-F))=1$.\label{footnote-F}}, the other way is using the fractionalization class of the 0-form symmetry $\Spin\times_{\Z_2^{\rF}}\frac{\U(1)_{X_n}}{\Z_n}$ for odd $n$ or $\Spin\times\frac{\U(1)_{X_n}}{\Z_n}$ for even $n$ and electric 1-form  symmetry $\Z_{6/{\rm q},[1]}^e$ of the gauged Standard Model to cancel the mixed anomaly of the electric 1-form symmetry $\Z_{6/{\rm q},[1]}^e$ and the magnetic 1-form symmetry $\U(1)_{[1]}^m$ of the Standard Model.

We first consider the first way, then
\bea\label{eq:current2-X}
J_m^{(2)}=\sigma_n\star \frac{\dd A_{X_n}}{2\pi}.
\eea
By \eqref{eq:constraint-X}, \eqref{eq:constraint-X-gauged}, \eqref{eq:current-X}, and \eqref{eq:current2-X}, we have 
\bea
\sigma_n=\frac{{\rm q}(1-n)\text{gcd}(2,n)}{2n}\mod 1.
\eea

Now we consider the second way. 
We consider the Standard Model Lagrangian. The Standard Model gauge group is $G_{\SM_{\rm q}}=\frac{\SU(3)\times \SU(2)\times \U(1)_{\tilde Y}}{\Z_{\rm q}}$. By turning on the background gauge field $B_e$ for the electric 1-form $\Z_{6/{\rm q},[1]}^e$ symmetry, 
the 2-form current $J_m^{(2)}$ becomes
\bea
J_m^{(2)}=\star \frac{{\rm q}\dd a_{\tilde{Y}}-\tilde{B}_e}{2\pi}
\eea
because it should be gauge invariant where $\tilde{B}_e=\frac{2\pi}{6/{\rm q}}B_e$, $\dd a_{\tilde Y}$ is the field strength of the dynamical gauge field $a_{\tilde Y}$ for the $\U(1)_{\tilde Y}$ symmetry. 
The coupling term $\int_{M^4}B_m\star J_m^{(2)}$ in the Standard Model Lagrangian becomes $\int_{M^4}B_m\frac{{\rm q}\dd a_{\tilde{Y}}-\tilde{B}_e}{2\pi}$ where $B_m$ is the background gauge field for the $\U(1)_{[1]}^m$ magnetic 1-form symmetry. However, this term is not gauge invariant under the gauge transformations
\bea
{\rm q}a_{\tilde Y}&\to&{\rm q}a_{\tilde Y}+\lambda,\nn\\
\tilde{B}_e&\to& \tilde{B}_e+\dd\lambda,\nn\\
B_m&\to&B_m+\dd\mu
\eea
where $\lambda=\frac{2\pi}{6/{\rm q}}C$ and $C$ is a $\Z_{6/{\rm q}}$-valued 1-cochain with the normalization $\oint \dd\lambda\in\frac{2\pi}{6/{\rm q}}\Z$, $\mu$ is a $\R$-valued 1-form with the normalization $\oint \dd\mu\in 2\pi\Z$.

Under the above gauge transformations,
\bea
B_m\frac{{\rm q}\dd a_{\tilde{Y}}-\tilde{B}_e}{2\pi}\to B_m\frac{{\rm q}\dd a_{\tilde{Y}}-\tilde{B}_e}{2\pi}+\dd \mu\frac{{\rm q}\dd a_{\tilde{Y}}-\tilde{B}_e}{2\pi}.
\eea
If $M^4$ is a closed 4-manifold which is the boundary of a 5-manifold $M^5$, then
\bea
\int_{M^4}\frac{{\rm q}\dd a_{\tilde{Y}}-\tilde{B}_e}{2\pi}\dd \mu&=&\int_{M^4}\frac{1}{2\pi}\dd(({\rm q}\dd a_{\tilde{Y}}-\tilde{B}_e) \mu)-\int_{M^4}\frac{1}{2\pi}\dd({\rm q}\dd a_{\tilde{Y}}-\tilde{B}_e) \mu\nn\\
&=&\int_{M^4}\frac{1}{2\pi}\dd\tilde{B}_e\mu\nn\\
&=&\int_{M^5}-\frac{1}{2\pi}\dd\tilde{B}_e\dd\mu.
\eea
There is an anomaly $\int_{M^5}\frac{1}{2\pi}B_m\dd \tilde{B}_e$ such that $\int_{M^4}\frac{1}{2\pi}({\rm q}\dd a_{\tilde{Y}}-\tilde{B}_e)B_m+\int_{M^5}\frac{1}{2\pi}B_m\dd \tilde{B}_e$ is gauge invariant. Combining $\int_{M^5}\frac{1}{2\pi}B_m\dd \tilde{B}_e$ and $\int_{M^4}-\frac{1}{2\pi}\tilde{B}_eB_m$, we get the anomaly 
\bea\label{eq:BedBm}
\int_{M^5}\frac{1}{2\pi}B_m\dd \tilde{B}_e + \int_{M^4}-\frac{1}{2\pi}\tilde{B}_eB_m
=
\int_{M^5}-\frac{1}{2\pi}\tilde{B}_e\dd B_m.
\eea

% Substituting the gauge bundle constraint \eqref{eq:constraint4-X} into the coupling term $\frac{1}{2\pi}(f_1-\frac{2\pi}{6}B_e)B_m$ in the Standard Model Lagrangian, we get the topological response 
% $\frac{1}{10}\frac{\dd A_X}{2\pi}B_m$ for $q=1$.

 The mixed anomaly of the electric $\Z_{6/{\rm q},[1]}^e$ 1-form symmetry and the magnetic $\U(1)_{[1]}^m$ 1-form symmetry of the Standard Model is $\int_{M^5}-\frac{1}{2\pi}\tilde{B}_e\dd B_m$ for ${\rm q}=1,2,3,6$ where $\tilde{B}_e=\frac{2\pi}{6/{\rm q}}B_e$. This anomaly appears as an SPTs or cobordism invariant in \Table{table:U1-q=3-TP} and \Table{table:U1-q=2-TP}.

In \Sec{sec:SF-odd} and \Sec{sec:SF-even}, we derive the constraint 
\bea
f_k^*\tilde{B}_e=\frac{k}{6/{\rm q}}\dd A_{X_n}
\eea
for the symmetry fractionalization of the 0-form $G_{[0]}=\Spin\times_{\Z_2^{\rF}}\frac{\U(1)_{X_n}}{\Z_n}$ symmetry for odd $n$ or the 0-form $G_{[0]}=\Spin\times\frac{\U(1)_{X_n}}{\Z_n}$ symmetry for even $n$ and the 1-form electric $G_{[1]}^e=\Z_{6/{\rm q},[1]}^e$ symmetry of the gauged Standard Model.
Here, $k\in\Z_{6/{\rm q}}=\H^2(\B G_{[0]},G_{[1]}^e)$ is represented by the map $f_k:\B G_{[0]}\to \B^2G_{[1]}^e$.

To cancel the anomaly $\int_{M^5}-\frac{1}{2\pi}\tilde{B}_e\dd B_m$, we need to add $\int_{M^4}\frac{1}{2\pi}\frac{k}{6/{\rm q}}B_m\dd A_{X_n}$ in the Standard Model Lagrangian.
 The total topological response is $\int_{M^4}\frac{\sigma_n}{2\pi}B_m\dd A_{X_n}$ where
 \bea
 \sigma_n=\frac{{\rm q}(1-n)\text{gcd}(2,n)}{2n}+\frac{k{\rm q}}{6}\mod1.
 \eea

\subsection{Gauge bundle constraint for odd $n$}\label{sec:GBC-odd}
For odd $n$, \eq{eq:constraint-X} becomes
\bea\label{eq:constraint-X-odd}
\frac{\dd a_{\tilde{Y}}}{2\pi}=\frac{1}{2 
 n}\frac{\dd A_{X_n}}{2\pi}-\frac{1}{2} w_2(TM) +\frac{1}{{\rm q}}w_2^{({\rm q})}\mod1.
\eea
In this subsection, we derive this gauge bundle constraint.

By \Table{table:SM}, there is a shared common subgroup $\Z_n$ between
$\U(1)_{ \tilde Y }$ and $\U(1)_{X_n}$, because the charge of $\mathbf{B}-\mathbf{L}-\frac{\tilde{Y}}{3}$ is an integer for all particles in \Table{table:SM}, hence the charges of $\U(1)_{ \tilde Y }$ and $\U(1)_{X_n}$ are equal modulo $n$ for all particles in \Table{table:SM}. 

For odd $n$, there is also a shared common subgroup $\Z_2^{\rF}$ between $\Spin$ and $\U(1)_{X_n}$; because the charge of $\mathbf{B}-\mathbf{L}-\frac{\tilde{Y}}{3}$ is an integer for all particles in \Table{table:SM}, hence the charges of $\U(1)_{X_n}$ and $\U(1)_{X_3=\mathbf{Q}-3\mathbf{L}}$ differ by
$(n-3)(\mathbf{B}-\mathbf{L}-\frac{\tilde{Y}}{3}) \in 2 \Z$, 
thus always have the same parity (even or odd) for all particles in \Table{table:SM} for odd $n$. Since the charge of $\U(1)_{X_3=\mathbf{Q}-3\mathbf{L}}$ is odd for all fermions in \Table{table:SM} and even for the Higgs boson in \Table{table:SM}, so is the charge of $\U(1)_{X_n}$ for odd $n$.

Since the symmetry should be faithful, 
% $\U(1)_{X}$ is from 
% $\Spin^c = \Spin \times_{\Z_2^\rF} \U(1)_{X}$.
the full spacetime-internal symmetry of the ungauged 
Standard Model is
\bea
{\Spin\times_{\Z_2^{\rm F}}\big(\frac{\U(1)_{ X_n }\times\frac{\SU(3)\times\SU(2)\times \U(1)_{ \tilde Y }}{\Z_{\rm q}}}{\Z_{n}} \big)}
\eea
for odd $n$.
Let $\tilde{w}_2^{(n)}$ be the obstruction class to lifting a $\frac{\frac{\SU(3)\times\SU(2)\times \U(1)_{ \tilde Y }}{\Z_{\rm q}}}{\Z_{n}}$ bundle to a $\frac{\SU(3)\times\SU(2)\times \U(1)_{ \tilde Y }}{\Z_{\rm q}}$ bundle, let 
$w_2^{({\rm q})}$ be the obstruction class to lifting a $\frac{\SU(3)\times\SU(2)}{\Z_{\rm q}}$ bundle to a $\SU(3)\times\SU(2)$ bundle, and 
let $\text{lcm}(n,{\rm q})c_{1,\tilde{Y}}$\footnote{ 
Consider the case when $\U(1) \supset {\Z_n}$ contains the $\Z_n$ as a normal subgroup.\\
$\bullet$ For the original $\U(1)$ with $c_1(\U(1))$, 
call the original U(1) gauge field $A$,
then $c_1=\frac{\dd A}{2 \pi} $.\\
$\bullet$ For the new $\U(1)'=\frac{\U(1)}{\Z_n}$ with $c_1(\U(1)')$,
call the new $\U(1)'$ gauge field $A'$,
then $c_1'=\frac{\dd A'}{2 \pi}=\frac{\dd (nA)}{2 \pi} = n c_1$.\\
$\bullet$ To explain why $A' = n A$ or $ c_1' = n c_1$, we look at the Wilson line operator
$\text{$\exp(\ii q' \oint A')$ and $\exp(\ii q \oint A)$.}$
The original $\U(1)$ has charge transformation $\exp(\ii q \theta)$ with $\theta \in [0, 2 \pi)$,
while the new $\U(1)'$ has charge transformation $\exp(\ii q' \theta')$ with $\theta' \in [0, 2 \pi)$.
But the $\U(1)'=\frac{\U(1)}{\Z_n}$, so the $\theta=\frac{2\pi}{n}$ in the old $\U(1)$ 
is identified as $\theta'=2\pi$ as a trivial zero
in the new $\U(1)'$.
In the original $\U(1)$, the $q \in \Z$ to be compatible with $\theta \in [0, 2 \pi)$.
In the new $\U(1)'$, the original $q$ is still allowed to have $n\Z$ to be compatible with $\theta \in [0, \frac{2\pi}{n})$;
but the new $q'=\frac{1}{n} q \in \Z$
and the new $\theta'= n \theta  \in [0, 2 \pi)$ are scaled accordingly.
Since the new $q'=\frac{1}{n} q \in \Z$, we show the new $A'=n A$.\label{footnote-c1}} (in terms of the first Chern class $c_{1,\tilde{Y}}$ of the $\U(1)_{\tilde{Y}}$ bundle) be the first Chern class of the $\frac{\U(1)_{\tilde{Y}}}{\Z_{\text{lcm}(n,{\rm q})}}$ bundle. By \cite{Benini_2017,Cheng_2023}, we have the following constraint
\bea\label{eq:constraint1-X-odd}
\text{lcm}(n,{\rm q})c_{1,\tilde{Y}}=\frac{{\rm q}}{\text{gcd}(n,{\rm q})}\tilde{w}_2^{(n)}+\frac{n}{\text{gcd}(n,{\rm q})}w_2^{({\rm q})}\mod \text{lcm}(n,{\rm q}).
\eea
Similarly, 
we also have the constraints
\bea\label{eq:constraint2-X-odd}
w_2(TM)&=&\mathfrak{o}^{(2)},\nn\\
2\tilde{w}_2^{(n)}+n \mathfrak{o}^{(2)}&=&2n c_{1,X_n }\mod 2n
\eea
where the $c_{1,X_n }$ is the first Chern class of the $\U(1)_{ X_n}$ bundle, 
the $2n c_{1,X_n }$ (see footnote \ref{footnote-c1}) is the first Chern class of the $\frac{\U(1)_{ X_n }}{\Z_{n}\times\Z_2^{\rF}}$ bundle, 
the $w_2(TM)$ is the second Stiefel-Whitney class of the tangent bundle which is the obstruction class to lifting a $\frac{\Spin}{\Z_2^{\rF}}$ bundle to a $\Spin$ bundle, and the $\mathfrak{o}^{(2)}$ is the obstruction class to lifting a $\frac{\frac{\U(1)_{X_n}\times \frac{\SU(3)\times\SU(2)\times \U(1)_{ \tilde Y }}{\Z_{\rm q}}}{\Z_n}}{\Z_2^{\rF}}$ bundle to a $\frac{\U(1)_{X_n}\times \frac{\SU(3)\times\SU(2)\times \U(1)_{ \tilde Y }}{\Z_{\rm q}}}{\Z_n}$ bundle.

Therefore, by \eqref{eq:constraint1-X-odd} and \eqref{eq:constraint2-X-odd}, we have
\bea
c_{1,\tilde{Y}}&=&\frac{\tilde{w}_2^{(n)}}{n}+\frac{w_2^{({\rm q})}}{{\rm q}}\mod 1\nn\\
&=&c_{1,X }-\frac{w_2(TM)}{2}+\frac{w_2^{({\rm q})}}{{\rm q}}\mod 1.
\eea

Let $a_{\tilde{Y}}$ be the gauge field for the $\U(1)_{\tilde{Y}}$ symmetry and $A_{X_n}$ the background gauge field for the $\frac{\U(1)_{ X_n }}{\Z_{n}\times\Z_2^{\rF}}$ symmetry. Then we have the gauge bundle constraint (see footnote \ref{footnote-c1})
\bea\label{eq:constraint4-X-odd}
\frac{\dd a_{\tilde{Y}}}{2\pi}=\frac{1}{2n}\frac{\dd A_{X_n}}{2\pi}-\frac{w_2(TM)}{2}+\frac{1}{{\rm q}}w_2^{({\rm q})}\mod1.
\eea

% By \cite{Gaiotto2014kfa1412.5148}, when the background gauge field $B_e$ for the electric 1-form symmetry $\Z_{6/q}$ of the Standard Model is not turned on, the $\U(1)$ magnetic 1-form symmetry is generated by a 2-form current 
% \bea\label{eq:current-X-odd}
% J_m^{(2)}= q\star \frac{\dd a_{\tilde{Y}}}{2\pi}.
% \eea
% Let $B_m$ be the background gauge field for the $\U(1)$ magnetic 1-form symmetry. 
% We consider the topological response
% $\int_{M^4} \frac{\sigma_n}{2\pi} B_m\dd A_{X_n}$ of the Standard Model, this term comes from the Standard Model Lagrangian in two ways. One way is substituting the gauge bundle constraint \eqref{eq:constraint4-X-odd} into the coupling term $B_m\star J_m^{(2)}$, the other way is using the fractionalization class of the 0-form $\frac{\U(1)_{ X_n }}{\Z_{n}\times\Z_2^{\rF}}$ symmetry and electric 1-form $\Z_{6/q}$ symmetry to cancel the mixed anomaly of the electric $\Z_{6/q}$ 1-form symmetry and the magnetic $\U(1)$ 1-form symmetry of the Standard Model.

% We first consider the first way, then
% \bea\label{eq:current2-X-odd}
% J_m^{(2)}=\sigma_n\star \frac{\dd A_{X_n}}{2\pi}.
% \eea
% By \eqref{eq:constraint4-X-odd}, \eqref{eq:current-X-odd}, and \eqref{eq:current2-X-odd}, we have 
% \bea
% \sigma_n=\frac{q}{2n}\mod 1
% \eea
% and we get another topological response $\int_{M^4}\frac{q}{2}w_2(TM)B_m$ of the Standard Model.

\subsection{Gauge bundle constraint for even $n$}\label{sec:GBC-even}
For even $n$, \eq{eq:constraint-X} becomes
\bea\label{eq:constraint-X-even}
\frac{\dd a_{\tilde{Y}}}{2\pi}=\frac{1}{ 
 n}\frac{\dd A_{X_n}}{2\pi} +\frac{1}{{\rm q}}w_2^{({\rm q})}\mod1.
\eea
In this subsection, we derive this gauge bundle constraint.

By \Table{table:SM}, there is a shared common subgroup $\Z_n$ between
$\U(1)_{ \tilde Y }$ and $\U(1)_{X_n}$, because the charge of $\mathbf{B}-\mathbf{L}-\frac{\tilde{Y}}{3}$ is an integer for all particles in \Table{table:SM}, hence the charges of $\U(1)_{ \tilde Y }$ and $\U(1)_{X_n}$ are equal modulo $n$ for all particles in \Table{table:SM}. 

For even $n$, there is no shared common subgroup $\Z_2^{\rF}$ between $\Spin$ and $\U(1)_{X_n}$; because the charge of $\mathbf{B}-\mathbf{L}-\frac{\tilde{Y}}{3}$ is an integer for all particles in \Table{table:SM}, hence the charges of $\U(1)_{X_n}$ and $\U(1)_{\tilde{Y}}$ differ by
$n(\mathbf{B}-\mathbf{L}-\frac{\tilde{Y}}{3}) \in 2 \Z$, 
thus always have the same parity (even or odd) for all particles in \Table{table:SM} for even $n$. Since the charge of $\U(1)_{\tilde{Y}}$ is not necessarily odd for all fermions in \Table{table:SM} and not even for the Higgs boson in \Table{table:SM}, so is the charge of $\U(1)_{X_n}$ for even $n$.

Since the symmetry should be faithful, 
% $\U(1)_{X}$ is from 
% $\Spin^c = \Spin \times_{\Z_2^\rF} \U(1)_{X}$.
the full spacetime-internal symmetry of the ungauged 
Standard Model is
\bea
{\Spin\times\big(\frac{\U(1)_{ X_n }\times\frac{\SU(3)\times\SU(2)\times \U(1)_{ \tilde Y }}{\Z_{\rm q}}}{\Z_{n}} \big)}
\eea
for even $n$.
Let $\tilde{w}_2^{(n)}$ be the obstruction class to lifting a $\frac{\frac{\SU(3)\times\SU(2)\times \U(1)_{ \tilde Y }}{\Z_{\rm q}}}{\Z_{n}}$ bundle to a $\frac{\SU(3)\times\SU(2)\times \U(1)_{ \tilde Y }}{\Z_{\rm q}}$ bundle, 
let $w_2^{({\rm q})}$ be the obstruction class to lifting a $\frac{\SU(3)\times\SU(2)}{\Z_{\rm q}}$ bundle to a $\SU(3)\times\SU(2)$ bundle, and 
let $\text{lcm}(n,{\rm q})c_{1,\tilde{Y}}$ (in terms of the first Chern class $c_{1,\tilde{Y}}$ of the $\U(1)_{\tilde{Y}}$ bundle) be the first Chern class of the $\frac{\U(1)_{\tilde{Y}}}{\Z_{\text{lcm}(n,{\rm q})}}$ bundle (see footnote \ref{footnote-c1}). By \cite{Benini_2017,Cheng_2023}, we have the following constraint
\bea\label{eq:constraint1-X-even}
\text{lcm}(n,{\rm q})c_{1,\tilde{Y}}=\frac{{\rm q}}{\text{gcd}(n,{\rm q})}\tilde{w}_2^{(n)}+\frac{n}{\text{gcd}(n,{\rm q})}w_2^{({\rm q})}\mod \text{lcm}(n,{\rm q}).
\eea
Similarly, 
we also have the constraints
\bea\label{eq:constraint2-X-even}
\tilde{w}_2^{(n)}=n c_{1,X_n }\mod n
\eea
where $c_{1,X }$ is the first Chern class of the $\U(1)_{ X_n}$ bundle, $n c_{1,X_n }$ (see footnote \ref{footnote-c1}) is the first Chern class of the $\frac{\U(1)_{ X_n }}{\Z_{n}}$ bundle.

Therefore, by \eqref{eq:constraint1-X-even} and \eqref{eq:constraint2-X-even}, we have
\bea
c_{1,\tilde{Y}}&=&\frac{\tilde{w}_2^{(n)}}{n}+\frac{w_2^{({\rm q})}}{{\rm q}}\mod 1\nn\\
&=&c_{1,X_n }+\frac{w_2^{({\rm q})}}{{\rm q}}\mod 1.
\eea

Let $a_{\tilde{Y}}$ be the gauge field for the $\U(1)_{\tilde{Y}}$ symmetry and $A_{X_n}$ the background gauge field for the $\frac{\U(1)_{ X_n }}{\Z_{n}}$ symmetry. Then we have the gauge bundle constraint (see footnote \ref{footnote-c1})
\bea\label{eq:constraint4-X-even}
\frac{\dd a_{\tilde{Y}}}{2\pi}=\frac{1}{n}\frac{\dd A_{X_n}}{2\pi}+\frac{1}{{\rm q}}w_2^{({\rm q})}\mod1.
\eea

\subsection{Symmetry Fractionalization for odd $n$}\label{sec:SF-odd}

% {\bf Lemma} The 0-form symmetry and 1-form symmetry are classified by $f_{[0]}:M\to\B G_{[0]}$ and $f_{[1]}:M\to\B^2 G_{[1]}$ respectively. The symmetry fractionalization is a map $g:\B G_{[0]}\to\B^2G_{[1]}$ such that $g\circ f_{[0]}=f_{[1]}$. The symmetry fractionalization class is the homotopy class of $g$ in $[\B G_{[0]},\B^2G_{[1]}]=\H^2(\B G_{[0]},G_{[1]})$.

In the presence of a higher form symmetry, a lower form symmetry requires a fractionalization class to completely specify its symmetry action on the physical theory, see also \App{app:fractionalization}. The fractionalization class of the 0-form $\Spin\times_{\Z_2^{\rF}}\frac{\U(1)_{ X_n }}{\Z_{n}}$ symmetry and electric 1-form $\Z_{6/{\rm q},[1]}^e$ symmetry of the gauged Standard Model is classified by 
\bea
k\in\H^2(\B(\Spin\times_{\Z_2^{\rF}}\frac{\U(1)_{ X_n }}{\Z_{n}}),\Z_{6/{\rm q}})=\Z_{6/{\rm q}}.
\eea
Since $\H^2(\B(\Spin\times_{\Z_2^{\rF}}\frac{\U(1)_{ X_n }}{\Z_{n}}),\Z_{6/{\rm q}})=[\B(\Spin\times_{\Z_2^{\rF}}\frac{\U(1)_{ X_n }}{\Z_{n}}),\B^2\Z_{6/{\rm q}}]$ is the group of homotopy classes of maps $\B(\Spin\times_{\Z_2^{\rF}}\frac{\U(1)_{ X_n }}{\Z_{n}})\to\B^2\Z_{6/{\rm q}}$. Let $k$ be represented by the map $f_k:\B(\Spin\times_{\Z_2^{\rF}}\frac{\U(1)_{ X_n }}{\Z_{n}})\to\B^2\Z_{6/{\rm q}}$. Since $B_e$ is the generator of $\H^2(\B^2\Z_{6/{\rm q}},\Z_{6/{\rm q}})$, and $\frac{\dd A_{X_n}}{2\pi}$ is the generator of $\H^2(\B(\Spin\times_{\Z_2^{\rF}}\frac{\U(1)_{ X_n }}{\Z_{n}}),\Z)=\Z$ (see \App{app:cohomology}),
the symmetry fractionalization requires the constraint (analogous to \cite{Hsin:2024lya2411.18160})
\bea
f_k^*B_e=k\frac{\dd A_{X_n}}{2\pi}
\eea
or equivalently
\bea
f_k^*\tilde{B}_e=\frac{k}{6/{\rm q}}\dd A_{X_n}
\eea
 with the normalization $\oint\tilde{B}_e\in \frac{2\pi}{6/{\rm q}}\Z$ where $\tilde{B}_e=\frac{2\pi}{6/{\rm q}}B_e$ and $\oint \dd A_{X_n}\in 2\pi\Z$.

 % To cancel the anomaly $\int_{M^5}-\frac{1}{2\pi}\tilde{B}_e\dd B_m$, we need to add $\int_{M^4}\frac{1}{2\pi}\frac{k}{6/q}B_m\dd A_{X_n}$ in the Standard Model Lagrangian.
 % The total topological response is $\int_{M^4}\frac{\sigma_n}{2\pi}B_m\dd A_{X_n}-\sigma_n'w_2(TM)B_m$ where
 % \bea
 % \sigma_n&=&\frac{q}{2n}+\frac{kq}{6}\mod1,\nn\\
 % \sigma_n'&=&\frac{q}{2}\mod1.
 % \eea

% \begin{enumerate}

% \item $\int_{M^4} B_m \star J_{m}^{(2)}$ and 
% $q\star \frac{\dd a_{\tilde{Y}}}{2\pi}$

% \item 
% $\int_{M^5}-\frac{1}{2\pi}\tilde{B}_e\dd B_m$
% \end{enumerate}

\subsection{Symmetry Fractionalization for even $n$}\label{sec:SF-even}

% {\bf Lemma} The 0-form symmetry and 1-form symmetry are classified by $f_{[0]}:M\to\B G_{[0]}$ and $f_{[1]}:M\to\B^2 G_{[1]}$ respectively. The symmetry fractionalization is a map $g:\B G_{[0]}\to\B^2G_{[1]}$ such that $g\circ f_{[0]}=f_{[1]}$. The symmetry fractionalization class is the homotopy class of $g$ in $[\B G_{[0]},\B^2G_{[1]}]=\H^2(\B G_{[0]},G_{[1]})$.

In the presence of a higher form symmetry, a lower form symmetry requires a fractionalization class to completely specify its symmetry action on the physical theory, see also \App{app:fractionalization}. The fractionalization class of the 0-form $\Spin\times\frac{\U(1)_{ X_n }}{\Z_{n}}$ symmetry and electric 1-form $\Z_{6/{\rm q},[1]}^e$ symmetry of the gauged Standard Model is classified by 
\bea
k\in\H^2(\B(\Spin\times\frac{\U(1)_{ X_n }}{\Z_{n}}),\Z_{6/{\rm q}})=\Z_{6/{\rm q}}.
\eea
Since $\H^2(\B(\Spin\times\frac{\U(1)_{ X_n }}{\Z_{n}}),\Z_{6/{\rm q}})=[\B(\Spin\times\frac{\U(1)_{ X_n }}{\Z_{n}}),\B^2\Z_{6/{\rm q}}]$ is the group of homotopy classes of maps $\B(\Spin\times\frac{\U(1)_{ X_n }}{\Z_{n}})\to\B^2\Z_{6/{\rm q}}$. Let $k$ be represented by the map $f_k:\B(\Spin\times\frac{\U(1)_{ X_n }}{\Z_{n}})\to\B^2\Z_{6/{\rm q}}$. Since $B_e$ is the generator of $\H^2(\B^2\Z_{6/{\rm q}},\Z_{6/{\rm q}})$, and $\frac{\dd A_{X_n}}{2\pi}$ is the generator of $\H^2(\B(\Spin\times\frac{\U(1)_{ X_n }}{\Z_{n}}),\Z)=\Z$ (see \App{app:cohomology}),
the symmetry fractionalization requires the constraint (analogous to \cite{Hsin:2024lya2411.18160})
\bea
f_k^*B_e=k\frac{\dd A_{X_n}}{2\pi}
\eea
or equivalently
\bea
f_k^*\tilde{B}_e=\frac{k}{6/{\rm q}}\dd A_{X_n}
\eea
 with the normalization $\oint\tilde{B}_e\in \frac{2\pi}{6/{\rm q}}\Z$ where $\tilde{B}_e=\frac{2\pi}{6/{\rm q}}B_e$ and $\oint \dd A_{X_n}\in 2\pi\Z$.

 % To cancel the anomaly $\int_{M^5}-\frac{1}{2\pi}\tilde{B}_e\dd B_m$, we need to add $\int_{M^4}\frac{1}{2\pi}\frac{k}{6/q}B_m\dd A_{X_n}$ in the Standard Model Lagrangian.
 % The total topological response is $\int_{M^4}\frac{\sigma_n}{2\pi}B_m\dd A_{X_n}$ where
 % \bea
 % \sigma_n=\frac{q}{n}+\frac{kq}{6}\mod1.
 % \eea

% \begin{enumerate}

% \item $\int_{M^4} B_m \star J_{m}^{(2)}$ and 
% $q\star \frac{\dd a_{\tilde{Y}}}{2\pi}$

% \item 
% $\int_{M^5}-\frac{1}{2\pi}\tilde{B}_e\dd B_m$
% \end{enumerate}

\section{Conclusion and Discussion:\\
Completely distinguish the Standard Model variants SM$_{({\rm q},k)}$ via $\sigma_n{({\rm q},k)}$
}\label{sec:conclusion}

\begin{table}[!h]
\begin{tabular}{| c |  c  | c | c |  c |  c |}
\hline\rule{0pt}{10pt}
$\sigma_n ({\rm q},k \in \Z_{6/{\rm q}})$ & ${\rm q}=1$ &  ${\rm q}=2$ & ${\rm q}=3$ & ${\rm q}=6$
 & discernible \\
\hline\rule{0pt}{10pt}
$n=1$ & $\left\{0,\frac{1}{6},\frac{1}{3},\frac{1}{2}, \frac{2}{3},\frac{5}{6}\right\}$ & $\left\{0,\frac{1}{3},\frac{2}{3}\right\}$ & 
$\left\{0,\frac{1}{2}\right\}$ & $\{0\}$ & No \\
\hline
$n=2$ & $\left\{\frac{1}{2},\frac{2}{3},\frac{5}{6},0,\frac{1}{6},\frac{1}{3}\right\}$ & $\left\{0,\frac{1}{3},\frac{2}{3}\right\}$ & $\left\{\frac{1}{2},0\right\}$ & $\{0\}$ & No\\
\hline
$n=3$ & $\left\{\frac{2}{3},\frac{5}{6},0,\frac{1}{6},\frac{1}{3},\frac{1}{2}\right\}$ &
$\left\{\frac{1}{3},\frac{2}{3},0\right\}$
& $\left\{0,\frac{1}{2}\right\}$ & $\{0\}$  
& 
{\parbox{1.8cm}{ 
\vspace{2pt}
No\\ 
\cite{Hsin:2024lya2411.18160}
\vspace{2pt}}}
\\
\hline
$n=4$ & 
$\left\{\frac{1}{4},\frac{5}{12},\frac{7}{12},\frac{3}{4},
\frac{11}{12},\frac{1}{12}\right\}$ 
& $\left\{\frac{1}{2},\frac{5}{6},\frac{1}{6}\right\}$ 
& $\left\{\frac{3}{4},\frac{1}{4}\right\}$ & $\left\{\frac{1}{2}\right\}$ & No\\
\hline
$n=5$ & 
$\left\{\frac{3}{5},\frac{23}{30},\frac{14}{15},\frac{1}{10},\frac{4}{15},\frac{13}{30}\right\}$ & 
$\left\{\frac{1}{5},\frac{8}{15},\frac{13}{15}\right\}$
& $\left\{\frac{4}{5},\frac{3}{10}\right\}$
& $\left\{\frac{3}{5}\right\}$ & No\\
\hline
$n=6$ & $\left\{\frac{1}{6},\frac{1}{3},\frac{1}{2},\frac{2}{3},\frac{5}{6},0\right\}$ & $\left\{\frac{1}{3},\frac{2}{3},0\right\}$ & $\left\{\frac{1}{2},0\right\}$ & $\{0\}$ & No\\
\hline
$n=7$ & $\left\{\frac{4}{7},\frac{31}{42},\frac{19}{21},\frac{1}{14},\frac{5}{21},
\frac{17}{42}\right\}$ & $\left\{\frac{1}{7},\frac{10}{21},\frac{17}{21}\right\}$ &
$\left\{\frac{5}{7},\frac{3}{14}\right\}$ & 
$\left\{\frac{3}{7}\right\}$ & Yes\\
\hline
$n=8$ & $\left\{\frac{1}{8},\frac{7}{24},\frac{11}{24},\frac{5}{8},\frac{19}{24},\frac{23}{24}\right\}$ & $\left\{\frac{1}{4},\frac{7}{12},\frac{11}{12}\right\}$ & $\left\{\frac{3}{8},\frac{7}{8}\right\}$ & $\left\{\frac{3}{4}\right\}$ & Yes\\
\hline
$n=9$ & $\left\{\frac{5}{9},\frac{13}{18},\frac{8}{9},\frac{1}{18},\frac{2}{9},\frac{7}{18}\right\}$ & $\left\{\frac{1}{9},\frac{4}{9},\frac{7}{9}\right\}$ & $\left\{\frac{2}{3},\frac{1}{6}\right\}$ & $\left\{\frac{1}{3}\right\}$ & Yes\\
\hline
$n=10$ & $\left\{\frac{1}{10},\frac{4}{15},\frac{13}{30},\frac{3}{5},\frac{23}{30},\frac{14}{15}\right\}$ & $\left\{\frac{1}{5},\frac{8}{15},\frac{13}{15}\right\}$ & $\left\{\frac{3}{10},\frac{4}{5}\right\}$ & $\left\{\frac{3}{5}\right\}$ & No\\
\hline
$n=11$ & $\left\{\frac{6}{11},\frac{47}{66},\frac{29}{33},\frac{1}{22},\frac{7}{33},\frac{25}{66}\right\}$ & $\left\{\frac{1}{11},\frac{14}{33},\frac{25}{33}\right\}$ & $\left\{\frac{7}{11},\frac{3}{22}\right\}$ & $\left\{\frac{3}{11}\right\}$ & Yes\\
\hline
$n=12$ & $\left\{\frac{1}{12},\frac{1}{4},\frac{5}{12},\frac{7}{12},\frac{3}{4},\frac{11}{12}\right\}$ & $\left\{\frac{1}{6},\frac{1}{2},\frac{5}{6}\right\}$ & $\left\{\frac{1}{4},\frac{3}{4}\right\}$ & $\left\{\frac{1}{2}\right\}$ & No\\
\hline
$\dots$ & $\dots$ & $\dots$ & $\dots$ & $\dots$ & $\dots$\\
\hline
\end{tabular}
\caption{Given a series of baryon minus lepton like global symmetry
$X_n \equiv 
n ({\bf B}-{\bf L}) + (1 - \frac{n}{3}) \tilde{Y}$ labeled by an integer $n\geq 1$,
we list down the numerical values of the 
fractional topological response
(similar to fractional
Hall coefficient)
$\sigma_n({\rm q}, k)=\frac{{\rm q}(1-n)\text{gcd}(2,n)}{2n}+\frac{k{\rm q}}{6}\mod1$
in the topological response action
$ \sigma_n({\rm q}, k) \int_{M^4}  \frac{1}{2\pi}{B_m\dd A_{X_n}}$ probed by 1-form background field $A_{X_n}$
and 2-form background field $B_m$, respectively background fields of 
$\U(1)_{X_n}$ 0-symmetry and
$\U(1)_{[1]}^m$ magnetic 1-symmetry.
Given an $n$ in the row,
we consider ${\rm q}=1,2,3,6$ in the column,
and the 
$\sigma_n ({\rm q},k \in \Z_{6/{\rm q}})$
values in the $\{..\}$ running from
$\{k=0,\dots, {6/{\rm q}}-1\}$ so $k \in \Z_{6/{\rm q}}$.
Here $n=3$ does not discern all the four ${\rm q}$ cases
\cite{Hsin:2024lya2411.18160}.
For a fixed $n$ specified by $X_n$,
the $\sigma_n$ can uniquely determine the
SM gauge group ${\rm q}=1,2,3,6$
and fractionalization class $k$ if and only if $n\ge7$ and $n\ne 10,12,15,30$.
Moreover, by choosing a pair of $n=2$ and $3$ together, then
we can further uniquely discern SM$_{({\rm q},k)}$ by measuring both of their $\sigma_2$ and $\sigma_3$. Similarly, pairs of
$\sigma_n$ with $n=(2,3)$, $(2,5)$, $(3,4)$, $(3,5)$, $(4,5)$, etc., all such 
pairs can discern SM$_{({\rm q},k)}$.
}
\label{table:response}
\end{table}

We derived the topological response $\int_{M^4}\frac{\sigma_n}{2\pi}B_m\dd A_{X_n}$ of the Standard Model where $A_{X_n}$ is the background gauge field for the $\frac{\U(1)_{ X_n }}{\Z_{\text{lcm}(2,n)}}$ symmetry, $B_m$ is the background gauge field for the $\U(1)_{[1]}^m$ magnetic 1-form symmetry, and
\bea
\sigma_n=\frac{{\rm q}(1-n)\text{gcd}(2,n)}{2n}+\frac{k{\rm q}}{6}\mod1.
 \eea

We discuss for which values of $n$, $\sigma_n$ can distinguish ${\rm q}=1,2,3,6$. 

For odd $n$, if $\sigma_n({\rm q},k)=\sigma_n({\rm q}',k')$, then
\bea
\frac{({\rm q}'-{\rm q})(n-1)}{2n}=\frac{k'{\rm q}'-k{\rm q}}{6}\mod1
\eea
or equivalently
\bea\label{eq:discussion-odd}
\frac{({\rm q}'-{\rm q})3(n-1)}{n}=k'{\rm q}'-k{\rm q}\mod6.
\eea
In particular, $\frac{({\rm q}'-{\rm q})3(n-1)}{n}\in\Z$. Since $\text{gcd}(n-1,n)=1$, we have $n\mid 3({\rm q}'-{\rm q})$. We assume that ${\rm q}'>{\rm q}$. Since ${\rm q}'-{\rm q}$ can only take values 1, 2, 3, 4, and 5, we have $n=1$, 3, 5, 9, or 15. 

If $n=1$, \eqref{eq:discussion-odd} becomes $0=k'{\rm q}'-k{\rm q}\mod6$. This equation has many solutions, for example, $k=0$ and $k'=0$. So $\sigma_1$ cannot distinguish ${\rm q}=1,2,3,6$.

If $n=3$, \eqref{eq:discussion-odd} becomes $2({\rm q}'-{\rm q})=k'{\rm q}'-k{\rm q}\mod6$ or equivalently $(k'-2){\rm q}'=(k-2){\rm q}\mod6$. This equation has at least one solution, for example, ${\rm q}=1$, $k=2$, ${\rm q}'=6$, and $k'=0$. So $\sigma_3$ cannot distinguish ${\rm q}=1,2,3,6$.

If $n=5$, \eqref{eq:discussion-odd} becomes $\frac{({\rm q}'-{\rm q})12}{5}=k'{\rm q}'-k{\rm q}\mod6$. The only possible solution is ${\rm q}'=6$, $k'=0$, ${\rm q}=1$, and $12=-k\mod6$ implies $k=0$. So $\sigma_5$ cannot distinguish ${\rm q}=1,2,3,6$.

If $n=9$, \eqref{eq:discussion-odd} becomes $\frac{({\rm q}'-{\rm q})8}{3}=k'{\rm q}'-k{\rm q}\mod6$. The only possible solution is ${\rm q}'=6$, $k'=0$, ${\rm q}=3$, and $8=-3k\mod6$ which is impossible. So $\sigma_9$ can distinguish ${\rm q}=1,2,3,6$.

If $n=15$, \eqref{eq:discussion-odd} becomes $\frac{({\rm q}'-{\rm q})14}{5}=k'{\rm q}'-k{\rm q}\mod6$. The only possible solution is ${\rm q}'=6$, $k'=0$, ${\rm q}=1$, and $14=-k\mod6$ implies $k=4$. So $\sigma_{15}$ cannot distinguish ${\rm q}=1,2,3,6$.

Therefore, $\sigma_n$ can distinguish ${\rm q}=1,2,3,6$ for all odd $n\ge7$ except $n=15$.

For even $n=2m$, if $\sigma_n({\rm q},k)=\sigma_n({\rm q}',k')$, then
\bea
\frac{({\rm q}-{\rm q}')}{n}=\frac{k'{\rm q}'-k{\rm q}}{6}\mod1
\eea
or equivalently
\bea\label{eq:discussion-even}
\frac{({\rm q}-{\rm q}')3}{m}=k'{\rm q}'-k{\rm q}\mod6.
\eea
In particular, $m\mid 3({\rm q}-{\rm q}')$. We assume that ${\rm q}'>{\rm q}$. Since ${\rm q}'-{\rm q}$ can only take values 1, 2, 3, 4, and 5, we have $m=1$, 2, 3, 4, 5, 6, 9, 12, or 15. 

If $m=1$, \eqref{eq:discussion-even} becomes $3({\rm q}-{\rm q}')=k'{\rm q}'-k{\rm q}\mod6$ or equivalently $(k+3){\rm q}=(k'+3){\rm q}'\mod6$. This equation has at least one solution, for example, ${\rm q}=1$, $k=3$, ${\rm q}'=6$, and $k'=0$. So $\sigma_2$ cannot distinguish ${\rm q}=1,2,3,6$.

If $m=2$, \eqref{eq:discussion-even} becomes $\frac{({\rm q}-{\rm q}')3}{2}=k'{\rm q}'-k{\rm q}\mod6$. This equation has at least one solution, for example, ${\rm q}=2$, $k=0$, ${\rm q}'=6$, and $k'=0$. So $\sigma_4$ cannot distinguish ${\rm q}=1,2,3,6$.

If $m=3$, \eqref{eq:discussion-even} becomes ${\rm q}-{\rm q}'=k'{\rm q}'-k{\rm q}\mod6$ or equivalently $(k+1){\rm q}=(k'+1){\rm q}'\mod6$. This equation has at least one solution, for example, ${\rm q}=1$, $k=5$, ${\rm q}'=6$, and $k'=0$. So $\sigma_6$ cannot distinguish ${\rm q}=1,2,3,6$.

If $m=4$, \eqref{eq:discussion-even} becomes $\frac{({\rm q}-{\rm q}')3}{4}=k'{\rm q}'-k{\rm q}\mod6$. The only possible solution is ${\rm q}'=6$, $k'=0$, ${\rm q}=2$, and $-3=-2k\mod6$ which is impossible. So $\sigma_8$ can distinguish ${\rm q}=1,2,3,6$.

If $m=5$, \eqref{eq:discussion-even} becomes $\frac{({\rm q}-{\rm q}')3}{5}=k'{\rm q}'-k{\rm q}\mod6$. The only possible solution is ${\rm q}'=6$, $k'=0$, ${\rm q}=1$, and $-3=-k\mod6$ implies $k=3$. So $\sigma_{10}$ cannot distinguish ${\rm q}=1,2,3,6$.

If $m=6$, \eqref{eq:discussion-even} becomes $\frac{{\rm q}-{\rm q}'}{2}=k'{\rm q}'-k{\rm q}\mod6$. This equation has at least one solution, for example, ${\rm q}=2$, $k=1$, ${\rm q}'=6$, and $k'=0$. So $\sigma_{12}$ cannot distinguish ${\rm q}=1,2,3,6$.

If $m=9$, \eqref{eq:discussion-even} becomes $\frac{{\rm q}-{\rm q}'}{3}=k'{\rm q}'-k{\rm q}\mod6$. The only possible solution is ${\rm q}'=6$, $k'=0$, ${\rm q}=3$, and $-1=-3k\mod6$ which is impossible. So $\sigma_{18}$ can distinguish ${\rm q}=1,2,3,6$.

If $m=12$, \eqref{eq:discussion-even} becomes $\frac{{\rm q}-{\rm q}'}{4}=k'{\rm q}'-k{\rm q}\mod6$. The only possible solution is ${\rm q}'=6$, $k'=0$, ${\rm q}=2$, and $-1=-2k\mod6$ which is impossible. So $\sigma_{24}$ can distinguish ${\rm q}=1,2,3,6$.

If $m=15$, \eqref{eq:discussion-even} becomes $\frac{{\rm q}-{\rm q}'}{5}=k'{\rm q}'-k{\rm q}\mod6$. The only possible solution is ${\rm q}'=6$, $k'=0$, ${\rm q}=1$, and $-1=-k\mod6$ implies $k=1$. So $\sigma_{30}$ cannot distinguish ${\rm q}=1,2,3,6$.

Therefore, $\sigma_n$ can distinguish ${\rm q}=1,2,3,6$ for all even $n\ge8$ except $n=10,12,30$.

We conclude that $\sigma_n$ can distinguish ${\rm q}=1,2,3,6$ if and only if $$
\text{$n\ge7$ and $n\ne 10,12,15,30$.}
$$
Moreover, by choosing a pair of $n=2$ and $3$ together, then
we can further uniquely discern SM$_{({\rm q},k)}$ by measuring both of their $\sigma_2$ and $\sigma_3$. Similarly, pairs of
$\sigma_{n_1}$ and $\sigma_{n_2}$ with 
$$
\text{$(n_1,n_2)=(2,3)$, $(2,5)$, $(3,4)$, $(3,5)$, $(4,5), \dots$, etc.,} 
$$ 
all such 
pairs can discern SM$_{({\rm q},k)}$, see \Table{table:response}.

 In \cite{Hsin:2024lya2411.18160}, they consider $\Spin\times \frac{\U(1)_{3(\mathbf{B} -\mathbf{L})}\times G_{\SM_{\rm q}}}{\Z_3}$ as the spacetime-internal symmetry of the Standard Model. Note that $3(\mathbf{B} -\mathbf{L})=X_3$ is a special case of our $X_n$. 
 However, the symmetry they consider is not faithful, because Spin and $\U(1)_{3(\mathbf{B} -\mathbf{L})}$ share a common $\Z_2^{\rF}$. Moreover, their result for the coefficient of the topological response of the Standard Model is $\xi({\rm q},k)=\frac{{\rm q}}{3}+\frac{k{\rm q}}{6}\mod 1$ for ${\rm q}=1,2,3,6$ and $k=0,1,\dots,\frac{6}{{\rm q}}-1$, while our result for the coefficients of the topological response of the Standard Model is
$\sigma_3({\rm q},k)=-\frac{{\rm q}}{3}+\frac{k{\rm q}}{6}\mod 1$. The difference between $\sigma_3$ and $\xi$ is due to different choices of the background field of the 0-form symmetry of the Standard Model: they choose the background field of the $\frac{\U(1)_{X_3}}{\Z_3}=\U(1)_{\mathbf{B} -\mathbf{L}}$, while we choose the background field of the $\frac{\U(1)_{X_3}}{\Z_3\times\Z_2^{\rF}}$. The two choices yield different coefficients of the topological response of the Standard Model coming from the gauge bundle constraint, while they yield the same coefficient of the topological response of the Standard Model coming from the symmetry fractionalization.
In comparison with \cite{Hsin:2024lya2411.18160},
Hsin-Gomis shows only the special case of our $n=3$ that cannot discern completely the four versions of SM$_{({\rm q},k)}$ with ${\rm q}=1,2,3,6$.
At $n=3$ in \Table{table:response},
\begin{enumerate}
    \item 
    $\sigma_3{({\rm 1},2)}=\sigma_3{({\rm 2},2)}=\sigma_3(3,0)=\sigma_3(6,0)=0$, so the topological response cannot distinguish SM$_{({\rm 1},2)}$, SM$_{({\rm 2},2)}$, SM$_{(3,0)}$, and SM$_{(6,0)}$.
    \item 
    $\sigma_3{({\rm 1},0)}=\sigma_3{({\rm 2},1)}=\frac{2}{3}$, so the topological response cannot distinguish SM$_{({\rm 1},0)}$ and SM$_{({\rm 2},1)}$.
    \item 
    $\sigma_3{({\rm 1},4)}=\sigma_3{({\rm 2},0)}=\frac{1}{3}$, so the topological response cannot distinguish SM$_{({\rm 1},4)}$ and SM$_{({\rm 2},0)}$.
    \item 
    $\sigma_3{({\rm 1},5)}=\sigma_3{({\rm 3},1)}=\frac{1}{2}$, so the topological response cannot distinguish SM$_{({\rm 1},5)}$ and SM$_{({\rm 3},1)}$.
\end{enumerate}
In summary, our introduction of $\U(1)_{X_n}$ symmetry
in \eq{eq:Xn} for generic $n$ in \Table{table:response}
extends the results of \cite{Hsin:2024lya2411.18160}, 
enabling a complete distinction of the SM variants
SM$_{({\rm q},k)}$ completely.

Here are some additional questions for future directions:
\begin{enumerate}
\item {\bf Investigating Potential Symmetry Extensions}:
There is a possible symmetry extension between the 1-form electric $G_{[1]}^e=\Z_{\frac{6}{{\rm q}},[1]}^e $ symmetry and 1-form magnetic $G_{[1]}^m=\U(1)_{[1]}^m$ symmetry. However, in this article, we write the 1-form symmetry of the SM as the direct product $G_{[1]}=G_{[1]}^e\times G_{[1]}^m$. 
We need to explore if this symmetry extension affects our results in this article.

    \item {\bf Baryon plus lepton and other discrete symmetries}: There are 
    discrete $({\bf B} + {\bf L})$ symmetries \cite{KorenProtonStability2204.01741,WangWanYou2204.08393,Wang:2025oow2502.21319}
    and possibly other symmetries survived in the SM. It will be beneficial
    to explore the topological response associated with those symmetries.
    
    \item {\bf Experimental Measurement of Topological Responses}:
    It is crucial to design and conduct experiments aimed at measuring the topological response $\sigma_n$, analogous to the quantum Hall conductance. Such empirical investigations will validate the theoretical predictions presented in this work and enhance our understanding of the SM's topological characteristics.
    
\end{enumerate}

\section{Acknowledgments}
We thank Yunqin Zheng for many helpful discussions, especially on crucially devising and illuminating the structure of the $X_n$ symmetry in \Sec{sec:Xn}.
ZW is supported by the NSFC Grant No. 12405001. JW is supported by Harvard University CMSA and 
LIMS fellow fund. YZY is supported by the National Science Foundation Grant No. DMR-2238360.\\

\appendix

\section{Mathematical Tools}

\subsection{Systematic Notations and Definitions}\label{app:notation}

A $p$-form symmetry $G$ on a manifold $M$ is characterized by a map $M\to \B^{p+1}G$ \cite{Gaiotto2014kfa1412.5148}. Throughout this article, all fields of a $p$-form symmetry $G$ (defined as a differential form or a cohomology class) are pulled back to $M$ via the classifying map $M\to \B^{p+1}G$.

% \bea
% A= A_{X_n} &\in& \Omega^1(\B\U(1),\R),\cr
% B_e &\in & \H^2(\B^2\Z_{\frac{6}{{\rm q}}},\Z_{\frac{6}{{\rm q}}}),\cr
% B_m&\in& \Omega^2(\B^2\U(1),\R)
% \eea
In this article, we use the following notations:
\begin{enumerate}
    \item 
$A_{X_n}\in\Omega^1(\B(\frac{\U(1)_{ X_n }}{\Z_{\text{lcm}(2,n)}}),\R)$ is the real-valued connection 1-form field for the $\frac{\U(1)_{ X_n }}{\Z_{\text{lcm}(2,n)}}$ symmetry,
$A=A_{X_n}$ for odd $n$,
\item 

$c_1=\frac{\dd A}{2 \pi}\in \H^2(\B(\frac{\U(1)_{ X_n }}{\Z_{\text{lcm}(2,n)}}),\Z)=\Z$ is the first Chern class of the $\frac{\U(1)_{ X_n }}{\Z_{\text{lcm}(2,n)}}$ bundle where $\dd$ is the differential operator, 

\item
$B_{e}\in \H^2(\B^2\Z_{\frac{6}{{\rm q}}},\Z_{\frac{6}{{\rm q}}})=\Z_{\frac{6}{{\rm q}}}$ is the 2-form field for the $\Z_{6/{\rm q},[1]}^e$ electric 1-form symmetry, while
$\tilde{B}_e=\frac{2\pi}{6/{\rm q}}B_e$.

\item

$P(B_{e,{\rm q}=3})$ is the Pontryagin square of $B_{e,{\rm q}=3}$ with $P(B_{e,{\rm q}=3})=B_{e,{\rm q}=3}^2\mod2$,

\item
$B_m\in \Omega^2(\B^2\U(1),\R)$ is the real-valued connection 2-form field for the $\U(1)_{[1]}^m$ magnetic 1-form symmetry, 
and $\frac{\dd B_m}{2\pi}\in\H^3(\B^2\U(1),\Z)=\Z$ is the generator of the cohomology group where $\dd$ is the differential operator,

\item
$w_i=w_i(TM)$ is the $i$-th Stiefel-Whitney class of the tangent bundle,

\item
$p_1$ is the first Pontryagin class of the tangent bundle,
\item
$\text{CS}_3^{TM}$ is the gravitational Chern-Simons 3-form for the first Pontryagin class $p_1$ of the tangent bundle,
\item
$w_2^{({\rm q})}$ is the obstruction class to lifting a $\frac{\SU(3)\times\SU(2)}{\Z_{\rm q}}$ bundle to a $\SU(3)\times\SU(2)$ bundle,
\item
$\tilde{w}_2^{(n)}$ is the obstruction class to lifting a $\frac{\frac{\SU(3)\times\SU(2)\times \U(1)_{ \tilde Y }}{\Z_{\rm q}}}{\Z_{n}}$ bundle to a $\frac{\SU(3)\times\SU(2)\times \U(1)_{ \tilde Y }}{\Z_{\rm q}}$ bundle,
\item
$c_{1,X_n }$ is the first Chern class of the $\U(1)_{ X_n}$ bundle,
\item
$a_{\tilde{Y}}\in\Omega^1(\B\U(1)_{\tilde{Y}},\R)$ is the real-valued connection 1-form field for the $\U(1)_{ \tilde{Y} }$ symmetry,
\item
$c_{1,\tilde{Y}}$ is the first Chern class of the $\U(1)_{ \tilde{Y}}$ bundle.

\end{enumerate}

We denote
$$
{\frac{{G_1} \times  {G_2 } }{{N_{\text{shared}}}}} \equiv {{G_1} \times_{{N_{\text{shared}}}}  {G_2 } }.
$$
The ${N_{\text{shared}}}$ is the shared common normal subgroup symmetry between ${G_1}$ and ${G_2 }$.

\subsection{Classifying Space}\label{app:classifying}

The classifying space \( \B G \) of a topological group \( G \) is a space such that:
\begin{enumerate}
    \item \( G \)-bundles over a space \( X \) are classified by homotopy classes of maps from \( X \) to \( \B G \).
   \[
   \{ G\text{-bundles over } X \} \cong [X, \B G].
   \]
   \item The space \( \B G \) has a principal \( G \)-bundle \( {\rm E} G \to \B G \), called the universal \( G \)-bundle, such that every \( G \)-bundle over any \( X \) is a pullback of \( {\rm E} G \to \B G \).
\end{enumerate}

The classifying space \( \B G \) is constructed as follows:
\begin{enumerate}
    \item \( {\rm E} G \): A contractible space on which \( G \) acts freely (e.g., \( G \) acting on itself by left translation).
    \item \( \B G = {\rm E} G / G \): The quotient of \( {\rm E} G \) by this free \( G \)-action.
\end{enumerate}

Here are some examples of the classifying space:
\begin{enumerate}
    \item For \( G = \U(1) \) (the circle group):
   \( {\rm E} G \) is the infinite-dimensional sphere \( S^\infty \) with \( \U(1) \)-action.
   \( \B G = \mathbb{CP}^\infty \), the infinite complex projective space.
  \item For \( G = \mathbb{Z}_n \) (finite cyclic group):
   \( \B G \) is the Eilenberg?MacLane space \( K(\mathbb{Z}_n, 1) \).
\end{enumerate}

\subsection{Higher Classifying Space}\label{app:higher}

A higher classifying space \( \B^n G \) generalizes \( \B G \) to higher homotopy contexts. It is used to classify \( n \)-form symmetries or other higher categorical structures:
 \( \B^n G \) is the classifying space for \( (n-1) \)-form \( G \)-gauge fields or \( G \)-symmetries.

The higher classifying space \( \B^n G \) is constructed as follows:
 For a topological group \( G \), recursively define higher classifying spaces:
  \[
  \B^n G = \B(\B^{n-1} G), \quad \B^0 G = G.
  \]
   \( \B^n G \) corresponds to the \( n \)-fold iterated classifying space of \( G \).

Here are some examples of the higher classifying space:
\begin{enumerate}
    \item \( \B^2 \U(1) \): The classifying space for 1-form \( \U(1) \) gauge fields.
    \item \( \B^n \mathbb{Z} \): The Eilenberg?MacLane space \( K(\mathbb{Z}, n) \), classifying \( n \)-dimensional cohomology with \( \mathbb{Z} \)-coefficients.
\end{enumerate}

The Eilenberg?MacLane space is related to cohomology as follows:
Maps from a space \( X \) to \( K(G,n) \) correspond to elements of the cohomology group \( \H^n(X, G) \):
\[
[X, K(G,n)] \cong \H^n(X, G).
\]
Here, $[X,Y]$ denotes the homotopy classes of maps from $X$ to $Y$.

Note that the higher classifying space $\B^n G$ and the Eilenberg?MacLane space $K(G,n)$ are not the same in general, they are the same only when $G$ is discrete. For example, $\B^n\U(1)=\B^{n+1}\Z=K(\Z,n+1)$. 

\subsection{Second and Third Integral Cohomology of $\B\Spin^c$ and $\B(\Spin \times \U(1))$ }\label{app:cohomology}
The group $\Spin^c$ is defined as $\frac{\Spin\times \U(1)}{\Z_2}$. The short exact sequence of groups 
\bea
1\to\U(1)\to\Spin^c\to\SO\to1
\eea
induces a fibration
\bea
\B\U(1)\to\B\Spin^c\to\B\SO.
\eea
We have the following Serre spectral sequence
\bea\label{eq:SSS}
\H^p(\B\SO,\H^q(\B\U(1),\Z))\Rightarrow \H^{p+q}(\B\Spin^c,\Z),
\eea
see \Fig{fig:SSS}.

\begin{figure}[!h]
\center
\begin{sseq}[grid=none,labelstep=1,entrysize=1.5cm]{0...5}{0...5}
\ssdrop{\Z}
\ssmoveto 1 0 
\ssdrop{0}
\ssmoveto 2 0
\ssdrop{0}
\ssmoveto 3 0
\ssdrop{\Z_2}
\ssmoveto 4 0
\ssdrop{\Z}
\ssmoveto 5 0
\ssdrop{\Z_2}

\ssmoveto 0 1
\ssdrop{0}
\ssmoveto 1 1
\ssdrop{0}
\ssmoveto 2 1
\ssdrop{0}
\ssmoveto 3 1
\ssdrop{0}
\ssmoveto 4 1
\ssdrop{0}
\ssmoveto 5 1
\ssdrop{0}

\ssmoveto 0 2
\ssdrop{\Z}
\ssmoveto 1 2
\ssdrop{0}
\ssmoveto 2 2
\ssdrop{0}
\ssmoveto 3 2
\ssdrop{\Z_2}
\ssmoveto 4 2
\ssdrop{\Z}
\ssmoveto 5 2
\ssdrop{\Z_2}

\ssmoveto 0 3
\ssdrop{0}
\ssmoveto 1 3
\ssdrop{0}
\ssmoveto 2 3
\ssdrop{0}
\ssmoveto 3 3
\ssdrop{0}
\ssmoveto 4 3
\ssdrop{0}
\ssmoveto 5 3
\ssdrop{0}

\ssmoveto 0 4
\ssdrop{\Z}
\ssmoveto 1 4
\ssdrop{0}
\ssmoveto 2 4
\ssdrop{0}
\ssmoveto 3 4
\ssdrop{\Z_2}
\ssmoveto 4 4
\ssdrop{\Z}
\ssmoveto 5 4
\ssdrop{\Z_2}

\ssmoveto 0 5
\ssdrop{0}
\ssmoveto 1 5
\ssdrop{0}
\ssmoveto 2 5
\ssdrop{0}
\ssmoveto 3 5
\ssdrop{0}
\ssmoveto 4 5
\ssdrop{0}
\ssmoveto 5 5
\ssdrop{0}

\ssmoveto 0 2
\ssarrow[color=black] 3 {-2}

\end{sseq}
\center
\caption{Serre spectral sequence \eqref{eq:SSS}. There is a nontrivial differential $d_3$ such that $d_3(c_1)=\beta w_2$, where $c_1$ is the first Chern class of the $\U(1)$ bundle, $w_2$ is the second Stiefel-Whitney class of the $\SO$ bundle, and $\beta:\H^2(-,\Z_2)\to\H^3(-,\Z)$ is the Bockstein homomorphism.}
\label{fig:SSS}
\end{figure}

Therefore, $\H^2(\B\Spin^c,\Z)=\Z$ and $\H^3(\B\Spin^c,\Z)=0$. The generator of $\H^2(\B\Spin^c,\Z)=\Z$ is $c_1'$ where $c_1'=2c_1$ is the first Chern class of the $\frac{\U(1)}{\Z_2}$ bundle. 

By the universal coefficient theorem relating cohomology with arbitrary coefficients to cohomology with $\Z$ coefficients, we have $\H^2(\B\Spin^c,\Z_{6/{\rm q}})=\Z_{6/{\rm q}}$.

Since 
\bea
\H^i(\B\Spin,\Z)=\left\{\begin{array}{lllll}
    \Z & i=0, \\
    0 & i=1, \\
    0 & i=2, \\
    0 & i=3, \\
    \Z & i=4,
\end{array}\right.
\eea
and 
\bea
\H^*(\B\U(1),\Z)=\Z[c_1]
\eea
where $c_1$ is the first Chern class of the $\U(1)$ bundle, by the K\"unneth theorem,
we have 
$\H^2(\B(\Spin\times\U(1)),\Z)=\Z$ and $\H^3(\B(\Spin\times\U(1)),\Z)=0$. The generator of $\H^2(\B(\Spin\times\U(1)),\Z)=\Z$ is $c_1$ where $c_1$ is the first Chern class of the $\U(1)$ bundle.

By the universal coefficient theorem relating cohomology with arbitrary coefficients to cohomology with $\Z$ coefficients, we have $\H^2(\B(\Spin\times\U(1)),\Z_{6/{\rm q}})=\Z_{6/{\rm q}}$.

\subsection{Symmetry-Enriched Standard Model SM$_{(\rm {\rm q},k)}$:\\
$(G_{[0]},\; 
G_{[1]},\; \rho,\; [\beta])$
and symmetry fractionalization 
$k$}

\label{sec:SETSM}

\subsubsection{$(G_{[0]},\; 
G_{[1]},\; \rho: \pi_1(\B G_{[0]}) \to {\rm Aut}(G_{[1]}),\; [\beta] \in  
[\B G_{[0]}, \B^3 G_{[1]}]_{\rho})$ and $k$}

\label{subsec:G0G1rhobeta}

We study the Symmetry-Enriched
SM labeled by the following data
\bea
(G_{[0]},G_{[1]},\rho,[\beta]).
\eea
In \cite{1803.09336} notation,
our $
(G_{[0]},\; 
G_{[1]},\; \rho,\; [\beta])$ is their 
$(G,\; \cA,\; \rho,\; [\beta])$.
Here the 0-symmetry is 
\bea \label{eq:G0}
G_{[0]}&=&\left\{\begin{array}{ll}\Spin\times_{\Z_2^{\rF}}\frac{\U(1)_{X_n}}{\Z_n}&\text{for odd }n,\\
\Spin\times\frac{\U(1)_{X_n}}{\Z_n}&\text{for even }n,
\end{array}\right.
\eea
the 1-symmetry is 
\bea
G_{[1]}=G_{[1]}^e \times G_{[1]}^m = \Z_{6/{\rm q},[1]}^e \times \U(1)_{[1]}^m. 
\eea

\begin{enumerate}

\item {\bf Homomorphism $\rho: \pi_1(\B G_{[0]}) \to {\rm Aut}(G_{[1]})$ as a twist} ---
In \cite{1803.09336}, their $\rho$ is a homomorphism $G\to {\rm Aut}(\cA)$. However, in the definition of twisted cohomology $\H^*_{\rho}(X,\cA)$, $\rho$ is a homomorphism 
$\pi_1(X)\to {\rm Aut}(\cA)$; so in our case $X=\B G_{[0]}$, $\rho$ is a homomorphism 
\bea
\rho: \pi_1(\B G_{[0]}) \to {\rm Aut}(G_{[1]}). 
\eea 
Only when $G_{[0]}$ is discrete, then 
$\pi_1(\B G_{[0]}) =\pi_0( G_{[0]})=G_{[0]}$. 

In our case, given the connected $G_{[0]}$ in \eq{eq:G0},
since $\pi_1(\B G_{[0]}) =\pi_0( G_{[0]})=0 $, 
 our specific $\rho: \pi_1(\B G_{[0]}) \to {\rm Aut}(G_{[1]})$ is trivial here. 
 We may denote $\rho=0$.

\item
{\bf Postnikov class $[\beta]$ as a twisted homotopy class}
\bea \label{eq:beta}
[\beta]\in[\B G_{[0]}, \B^3 G_{[1]}]_{\rho},
\eea
which is
equivalent to (\emph{if and only if} $G_{[1]}$ is discrete) a degree-3 twisted cohomology class \cite{Barkeshli:2014cna1410.4540,1803.09336}, 
$$\H^3_{\rho}(\B G_{[0]},G_{[1]} ).$$

Regarding $\rho$:
\begin{enumerate}

\item 
{\bf When $\rho$ is nontrivial}, the Postnikov class $[\beta]$ is a twisted homotopy class or cohomology class.

\item 
{\bf When $\rho$ is trivial} (our case), the Postnikov class $[\beta]$ becomes an ordinary homotopy class or cohomology class.
Due to the connected $G_{[0]}$ in \eq{eq:G0}, we have derived already our $\rho=0$ is indeed trivial in this present work.

\end{enumerate}

Regarding the electric and magnetic sectors:
\begin{enumerate}
\item 
Electric
$[\beta^e]\in[\B G_{[0]}, \B^3 G_{[1]}^e]_{\rho}$:

In the case $G_{[1]}^e= \Z_{6/{\rm q}}$ and 
$\B^3G_{[1]}^e=
\B^3 \Z_{6/{\rm q}}= K(\Z_{6/{\rm q}},3)$ as the Eilenberg-MacLane space, the Postnikov class $[\beta^e]$ is a degree-3 twisted cohomology class 
$$\H^3_{\rho}(\B G_{[0]}, \Z_{6/{\rm q}} ),$$ 
because the degree-$l$ 
cohomology class with coefficients in discrete $\cA$ is a homotopy class of maps to the Eilenberg-MacLane space,
$K(\cA,l)$, here $\cA=\Z_{6/{\rm q}}$ and $l=3$.

By the results in \App{app:cohomology} and the universal coefficient theorem relating cohomology with arbitrary coefficients (here our $\Z_{6/{\rm q}}$) to cohomology with $\Z$ coefficients, the electric Postnikov class 
\bea \label{eq:betae}
[\beta^e]\in[\B G_{[0]}, \B^3 G_{[1]}^e]_{\rho}=\H^3_{\rho}(\B G_{[0]},G_{[1]}^e)=0
\eea is trivial, namely $[\beta^e]=0$.
Because $[\beta^e] = 0$,
we can define the 
symmetry fractionalization class $k^e$ in the electric sector.

\item 
Magnetic
$[\beta^m]\in[\B G_{[0]}, \B^3 G_{[1]}^m]_{\rho}$:

In the case $G_{[1]}^m=\U(1)$ and $\B^3G_{[1]}^m=
\B^3 \U(1)= K(\Z,4)$ as the Eilenberg-MacLane space, the Postnikov class $[\beta^m]$ is a degree-4 twisted cohomology class 
$$\H^4_{\rho}(\B G_{[0]},\Z ),$$ 
because the degree-$l$ 
cohomology class with coefficients in discrete $\cA$ is a homotopy class of maps to the Eilenberg-MacLane space,
$K(\cA,l)$, here $\cA=\Z$ and $l=4$.

The magnetic Postnikov class 
\bea \label{eq:betam}
[\beta^m] 
\in [\B G_{[0]}, \B^3 G_{[1]}^m]_{\rho} = 
\H^4_{\rho}(\B G_{[0]}, \Z)=\Z^2
\eea
may be a nontrivial obstruction class.
If $[\beta^m] \neq 0$ is nontrivial,
we will not be able to define the 
symmetry fractionalization class $k^m$ in the magnetic sector later.

\end{enumerate}

\item {\bf Symmetry fractionalization class $k$} \cite{Barkeshli:2014cna1410.4540,1803.09336}:
\bea
k \in  
[\B G_{[0]}, \B^2 G_{[1]}]_{\rho}.
\eea
Only when the earlier Postnikov class $[\beta]$ is trivial in
\eq{eq:beta},
namely $[\beta] = 0 \in[\B G_{[0]}, \B^3 G_{[1]}]_{\rho}$,
the symmetry fractionalization class $k$ can be defined.

\begin{enumerate}
\item 
Electric
$k^e\in[\B G_{[0]}, \B^2 G_{[1]}^e]_{\rho}$:

Thus, the symmetry fractionalization
in the electric sector
\bea
k^e \in  
[\B G_{[0]}, \B^2 G_{[1]}^e]_{\rho}
=
\H^2_{\rho}(\B G_{[0]}, G_{[1]}^e)
= \Z_{6/{\rm q}}
\eea
can be defined, because the obstruction $[\beta^e]=0$.

\item 
Magnetic
$k^m\in[\B G_{[0]}, \B^2 G_{[1]}^m]_{\rho}$:

\begin{itemize}

\item If $[\beta^m] \neq 0$ is nontrivial in \eq{eq:betam},
the symmetry fractionalization is not defined.

\item If $[\beta^m]=0$ is trivial in \eq{eq:betam}, 
by the results in \App{app:cohomology}, the symmetry fractionalization
in the magnetic sector $k^m\in [\B G_{[0]}, \B^2 G_{[1]}^m]_{\rho}
=
\H^3_{\rho}(\B G_{[0]}, \Z)
= 0$
is trivial.

\end{itemize}

\end{enumerate}

\end{enumerate}

\subsubsection{Explain the Symmetry fractionalization 
$k \in  
[\B G_{[0]}, \B^2 G_{[1]}]_{\rho}$ when $[\beta]=0$ 
via a lifting diagram}
\label{app:fractionalization}

In the presence of a higher form symmetry, a lower form symmetry requires a fractionalization class to completely specify its symmetry action on the physical theory. 
In this subsection, we consider that homomorphism  of the previous section $\rho=0$, which is the case of the SM study here.

We consider the following symmetry extension of the 0-form symmetry $G_{[0]}$ by the 1-form symmetry $G_{[1]}$
in terms of $\B\mathbb{G}$ fibration over $\B G_{[0]}$ with fiber $\B^2G_{[1]}$:
\bea\label{eq:symmetry-extension}
\B^2G_{[1]}\to \B\mathbb{G} \to \B G_{[0]}
\eea
where $\mathbb{G}$ is the total symmetry.
Since \eqref{eq:symmetry-extension} is a fibration with cofiber $\B^3G_{[1]}$, it induces a long exact sequence of groups
\bea
\cdots\to[\B G_{[0]},\B^2G_{[1]}]\to[\B G_{[0]},\B\mathbb{G}]\to [\B G_{[0]},\B G_{[0]}]\to[\B G_{[0]},\B^3G_{[1]}]\to\cdots.
\eea
So the lifting (dashed arrow) in the following diagram 
\bea\label{eq:lifting}
\xymatrix{\B^2G_{[1]}\ar[r]&\B\mathbb{G}\ar[d]&\\
\B G_{[0]}\ar[r]^{\text{id}}\ar@{-->}[ru]&\B G_{[0]} \ar[r]^{[\beta]}&\B^3G_{[1]}}
\eea
exists if and only if $[\beta]=0$ where $[\beta]$ is the Postnikov class classifying the symmetry extension \eqref{eq:symmetry-extension}.
The set of liftings in the diagram \eqref{eq:lifting} is a $[\B G_{[0]},\B^2 G_{[1]}]$-torsor, namely, $[\B G_{[0]},\B^2 G_{[1]}]$ acts on the set of liftings in the diagram \eqref{eq:lifting} simply transitively.

We call such a lifting in the diagram \eqref{eq:lifting} a symmetry fractionalization of the 0-form symmetry $G_{[0]}$ and the 1-form symmetry $G_{[1]}$, and the symmetry fractionalization is classified by a homotopy class 
\bea
k\in[\B G_{[0]},\B^2 G_{[1]}].
\eea 
The obstruction class to the symmetry fractionalization is 
\bea
[\beta]\in [\B G_{[0]},\B^3 G_{[1]}].
\eea

The 0-form symmetry and 1-form symmetry are classified by 
\bea
f_{[0]}:M\to\B G_{[0]}
\eea
and 
\bea
f_{[1]}:M\to\B^2 G_{[1]}
\eea 
respectively. The symmetry fractionalization is a map 
\bea
g:\B G_{[0]}\to\B^2G_{[1]}
\eea 
such that 
\bea
g\circ f_{[0]}=f_{[1]}.
\eea
The symmetry fractionalization class is the homotopy class of $g$ in 
\bea
[\B G_{[0]},\B^2G_{[1]}]=\H^2(\B G_{[0]},G_{[1]}),
\eea 
where identity holds \emph{if and only if} $G_{[1]}$ is discrete.

To connect back to the Standard Model (SM) example in the main text and in \App{subsec:G0G1rhobeta}, we summarize again the symmetry fractionalization in the SM as follows: 

\begin{itemize}

\item 
For odd $n$, there is a symmetry fractionalization of the 0-form $\Spin\times_{\Z_2^{\rF}}\frac{\U(1)_{ X_n }}{\Z_{n}}$ symmetry and electric 1-form $\Z_{6/{\rm q},[1]}^e$ symmetry of the gauged SM because $[\B (\Spin\times_{\Z_2^{\rF}}\frac{\U(1)_{ X_n }}{\Z_{n}}),\B^2\Z_{6/{\rm q}}]=
\H^2(\B (\Spin\times_{\Z_2^{\rF}}\frac{\U(1)_{ X_n }}{\Z_{n}}),\Z_{6/{\rm q}})=\Z_{6/{\rm q}}$ is nonzero (see \App{app:cohomology}).

However, there is no symmetry fractionalization of the 0-form $\Spin\times_{\Z_2^{\rF}}\frac{\U(1)_{ X_n }}{\Z_{n}}$ symmetry and magnetic 1-form $\U(1)_{[1]}^m$ symmetry of the gauged SM because $[\B (\Spin\times_{\Z_2^{\rF}}\frac{\U(1)_{ X_n }}{\Z_{n}}),\B^2\U(1)]=\H^3(\B (\Spin\times_{\Z_2^{\rF}}\frac{\U(1)_{ X_n }}{\Z_{n}}),\Z)=0$ (see \App{app:cohomology}).

\item 
Similarly, for even $n$, there is a symmetry fractionalization of the 0-form $\Spin\times\frac{\U(1)_{ X_n }}{\Z_{n}}$ symmetry and electric 1-form $\Z_{6/{\rm q},[1]}^e$ symmetry of the gauged SM because $\H^2(\B (\Spin\times\frac{\U(1)_{ X_n }}{\Z_{n}}),\Z_{6/{\rm q}})=\Z_{6/{\rm q}}$ is nonzero (see \App{app:cohomology}).
 
However, there is no symmetry fractionalization of the 0-form $\Spin\times\frac{\U(1)_{ X_n }}{\Z_{n}}$ symmetry and magnetic 1-form $\U(1)_{[1]}^m$ symmetry of the gauged SM because $[\B (\Spin\times\frac{\U(1)_{ X_n }}{\Z_{n}}),\B^2\U(1)]=\H^3(\B (\Spin\times\frac{\U(1)_{ X_n }}{\Z_{n}}),\Z)=0$ (see \App{app:cohomology}).

\end{itemize}

\subsection{Cobordism group classification}\label{sec:cobordism}

In this appendix, we review the notions of bordism groups and TP groups, see \cite{WanWang2018bns1812.11967} for a comprehensive introduction.
In addition, we compute new bordism groups and TP groups that are related to the Standard Model (SM) with 1-form symmetries. In \cite{WangWanYou2112.14765, WangWanYou2204.08393}, we compute the 
SM bordism groups $\Omega_d^{\Spin^c\times\B\Z_{6/{\rm q}}\times\B\U(1)}$ and the TP groups $\TP_d(\Spin^c\times\B\Z_{6/{\rm q}}\times\B\U(1))$ for  $d=5$ and ${\rm q}=1,2,3,6$. 
Here we expand the results in \cite{WangWanYou2112.14765, WangWanYou2204.08393},
including all dimensions $d=0,1,2,3,4,5$.

The \( d \)-dimensional bordism group \( \Omega_d^G \) is the group of equivalence classes of \( d \)-dimensional manifolds equipped with tangential \( G \)-structure, where two \( G \)-manifolds are considered equivalent if they can be the boundaries of a \( (d+1) \)-dimensional \( G \)-manifold.
 Bordism is an equivalence relation:
   \( M_0 \sim M_1 \) if there exists a \( (d+1) \)-dimensional \( G \)-manifold \( W \) with \( \partial W = M_0 \sqcup (-M_1) \).
 \( \Omega_d^G \) can be defined as the set of bordism classes of \( d \)-dimensional \( G \)-manifolds, equipped with an abelian group structure given by disjoint union:
  \[
  [M_0] + [M_1] = [M_0 \sqcup M_1].
  \]

The topological phases group \cite{Freed2016} \( \text{TP}_d(G) := [\text{MT}G, \Sigma^{d+1} I\mathbb{Z}(1)] \) classifies deformation-equivalent topological phases of quantum systems protected by a symmetry group \( G \) in \( d \)-dimensions.

Here?s a breakdown of the notation:
\begin{enumerate}
    \item \( \text{MT}G \):
    This is the Madsen-Tillmann spectrum associated with the \( G \)-structure. By the Pontryagin-Thom isomorphism, the bordism group $\Omega_d^G=\pi_d\text{MT}G$ is the homotopy group of $\text{MT}G$.
   \item \( \Sigma^{d+1} I\mathbb{Z}(1) \):
    This represents a $(d+1)$-fold suspension of \( I\mathbb{Z}(1) \), the Anderson dual to the sphere spectrum.
   \item \( [X, Y] \):
    This denotes the set of homotopy classes of maps between spectra \( X \) and \( Y \).
\end{enumerate}

\( \text{TP}_d(G) \) is computed as the space of all invertible field theories (or invertible topological phases) in \( d \)-dimensions, protected by the symmetry \( G \).
 The group structure on \( \text{TP}_d(G) \) corresponds to stacking topological phases.

The classification of invertible topological phases is deeply connected to bordism. There is a split short exact sequence of groups:
\[
0\to \Ext^1(\Omega_d^G,\Z)\to\text{TP}_d(G) \to\text{Hom}(\Omega_{d+1}^G, \mathbb{Z})\to0.
\]
This means:
\begin{enumerate}
    \item The torsion part of $\TP_d(G)$ is the same as the torsion part of $\Omega_d^G$.
    \item The free part of $\TP_d(G)$ is the same as the free part of $\Omega_{d+1}^G$.
\end{enumerate}

In the following \Table{table:U1-q=3-bordism}, \Table{table:U1-q=3-TP}, \Table{table:U1-q=2-bordism}, and \Table{table:U1-q=2-TP}, we compute the bordism groups $\Omega_d^{\Spin^c\times\B\Z_{6/{\rm q}}\times\B\U(1)}$ and the TP groups $\TP_d(\Spin^c\times\B\Z_{6/{\rm q}}\times\B\U(1))$ for ${\rm q}=2,3$ and $d\le5$.
We also list down bordism and cobordism invariants.  The results for ${\rm q}=1$ can be obtained by combining the results in \Table{table:U1-q=3-bordism}, \Table{table:U1-q=3-TP}, \Table{table:U1-q=2-bordism}, and \Table{table:U1-q=2-TP}. While the results for ${\rm q}=6$ can be obtained by removing the torsion classes from either \Table{table:U1-q=3-bordism} and \Table{table:U1-q=3-TP} or \Table{table:U1-q=2-bordism} and \Table{table:U1-q=2-TP}.

\begin{table}[!h]
\begin{center}
\begin{tabular}{c|c|c}

$d$ & $\Omega_d^{\Spin\times_{\Z_2}\U(1)\times\B\U(1)\times\B\Z_2}$ & bordism invariants\\
\hline
0 & $\Z$ \\
1 & $0$ \\
2 & $\Z\times\Z_2$ & $\frac{c_1}{2},B_{e,{\rm q}=3}$ \\
3 & $\Z$ & $\dd B_m$ \\
4 & $\Z^2\times\Z_4$  & $c_1^2,\frac{p_1}{24},P(B_{e,{\rm q}=3})$\\
5 & $\Z\times\Z_2$ & $c_1\dd B_m, B_{e,{\rm q}=3}\dd B_m$ \\
6 & $\Z^2\times\Z_2^3$ & $\frac{c_1^3}{2},\frac{c_1^3-c_1p_1}{48},B_{e,{\rm q}=3}^3,c_1^2B_{e,{\rm q}=3},\frac{c_1P(B_{e,{\rm q}=3})}{2}$\\
\end{tabular}
\caption{$\Spin\times_{\Z_2}\U(1)\times\B\U(1)\times\B\Z_2$ bordism groups for ${\rm q}=3$. We use the notations in \App{app:notation}. Here, $c_1=w_2=0\mod2$ on oriented 2-manifolds because $w_2+w_1^2=0$ on 2-manifolds. By the Wu formula, $c_1^3=w_2c_1^2=\Sq^2c_1^2=0\mod2$ on oriented 6-manifolds. $\frac{p_1}{24}$ and $\frac{c_1^3-c_1p_1}{48}$ are integer-valued by the Atiyah-Singer index theorem. By the Wu formula, $c_1P(B_{e,{\rm q}=3})=w_2B_{e,{\rm q}=3}^2=\Sq^2B_{e,{\rm q}=3}^2=0\mod2$ on oriented 6-manifolds. }
\label{table:U1-q=3-bordism}
\end{center}
\end{table} 

\begin{table}[!h]
\begin{center}
\begin{tabular}{c|c|c}

$d$ & $\TP_d(\Spin\times_{\Z_2}\U(1)\times\B\U(1)\times\B\Z_2)$ & cobordism invariants\\
\hline
0 & $0$ \\
1 & $\Z$  & $\frac{A}{2}$\\
2 & $\Z\times\Z_2$ & $B_m,B_{e,{\rm q}=3}$ \\
3 & $\Z^2$ & $A\dd A,\frac{\text{CS}_3^{TM}}{24}$\\
4 & $\Z\times\Z_4$ & $A\dd B_m,P(B_{e,{\rm q}=3})$ \\
5 & $\Z^2\times\Z_2$ & $\frac{A\dd A\dd A}{2},\frac{A\dd A\dd A-Ap_1}{48}, B_{e,{\rm q}=3}\dd B_m$\\
\end{tabular}
\caption{$\Spin\times_{\Z_2}\U(1)\times\B\U(1)\times\B\Z_2$ TP groups for ${\rm q}=3$. We use the notations in \App{app:notation}. The TP groups in this table are obtained by shifting the free $\Z$ classes in \Table{table:U1-q=3-bordism} to the one lower dimension while keeping the torsion classes. The cobordism invariants in this table are obtained by removing one differential operator $\dd$ in the bordism invariants in \Table{table:U1-q=3-bordism} for free $\Z$ classes while keeping the bordism invariants in \Table{table:U1-q=3-bordism} for torsion classes. We omit some $\frac{1}{2\pi}$ factors here. The term $\int_{M^4} \frac{1}{2\pi} B_m\dd A_{X_n}=\int_{M^4}\frac{1}{2\pi}A_{X_n}\dd B_m$ in \eqref{eq:BmdA} and the anomaly \eqref{eq:BedBm} appear as SPTs or cobordism invariants here.}
\label{table:U1-q=3-TP}
\end{center}
\end{table}

\begin{table}[!h]
\begin{center}
\begin{tabular}{c|c|c}

$d$ & $\Omega_d^{\Spin\times_{\Z_2}\U(1)\times\B\U(1)\times\B\Z_3}$ & bordism invariants\\
\hline
0 & $\Z$ \\
1 & $0$ \\
2 & $\Z\times\Z_3$ & $\frac{c_1}{2}, B_{e,{\rm q}=2}$\\
3 & $\Z$ &$\dd B_m$\\
4 & $\Z^2\times\Z_3^2$ &$c_1^2,\frac{p_1}{24}, B_{e,{\rm q}=2}^2,c_1B_{e,{\rm q}=2}$ \\
5 & $\Z\times\Z_3$ & $c_1\dd B_m,B_{e,{\rm q}=2}\dd B_m$\\
\end{tabular}
\caption{$\Spin\times_{\Z_2}\U(1)\times\B\U(1)\times\B\Z_3$ bordism groups for ${\rm q}=2$. We use the notations in \App{app:notation}. Similar explanations as in \Table{table:U1-q=3-bordism} apply here. We do not compute the 6d bordism group but we know that there are two free $\Z$ classes in the 6d bordism group similarly as in \Table{table:U1-q=3-bordism}.}
\label{table:U1-q=2-bordism}
\end{center}
\end{table} 

\begin{table}[!h]
\begin{center}
\begin{tabular}{c|c|c}

$d$ & $\TP_d(\Spin\times_{\Z_2}\U(1)\times\B\U(1)\times\B\Z_3)$ & cobordism invariants\\
\hline
0 & $0$ \\
1 & $\Z$ & $\frac{A}{2}$ \\
2 & $\Z\times\Z_3$ & $B_m,B_{e,{\rm q}=2}$\\
3 & $\Z^2$ & $A\dd A,\frac{\text{CS}_3^{TM}}{24}$\\
4 & $\Z\times\Z_3^2$ & $A\dd B_m, B_{e,{\rm q}=2}^2,c_1B_{e,{\rm q}=2}$ \\
5 & $\Z^2\times\Z_3$ & $\frac{A\dd A\dd A}{2},\frac{A\dd A\dd A-Ap_1}{48}, B_{e,{\rm q}=2}\dd B_m$\\
\end{tabular}
\caption{$\Spin\times_{\Z_2}\U(1)\times\B\U(1)\times\B\Z_3$ TP groups for ${\rm q}=2$. We use the notations in \App{app:notation}. The TP groups in this table are obtained by shifting the free $\Z$ classes in \Table{table:U1-q=2-bordism} to the one lower dimension while keeping the torsion classes. The cobordism invariants in this table are obtained by removing one differential operator $\dd$ in the bordism invariants in \Table{table:U1-q=2-bordism} for free $\Z$ classes while keeping the bordism invariants in \Table{table:U1-q=2-bordism} for torsion classes. We omit some $\frac{1}{2\pi}$ factors here. The term $\int_{M^4} \frac{1}{2\pi} B_m\dd A_{X_n}=\int_{M^4}\frac{1}{2\pi}A_{X_n}\dd B_m$ in \eqref{eq:BmdA} and the anomaly \eqref{eq:BedBm} appear as SPTs or cobordism invariants here.}
\label{table:U1-q=2-TP}
\end{center}
\end{table} 

%\twocolumngrid

%\tableofcontents

\onecolumngrid

\newpage

\bibliography{BSM-Proton.bib}

\end{document}